# The Orientations of the Giza pyramids and associated structures

*Erin Nell and Clive Ruggles*

School of Archaeology and Ancient History, University of Leicester,
Leicester LE8 0PJ, United Kingdom

## Long abstract

Ever since Flinders Petrie undertook a theodolite survey on the Giza plateau in 1881 and drew attention to the extraordinary degree of precision with which the three colossal pyramids are oriented upon the four cardinal directions, there have been a great many suggestions as to how this was achieved and why it was of importance. Surprisingly, perhaps—especially in view of the various interpretations that have been offered and the many other astronomical hypotheses and speculations, both scholarly and popular, that have been offered in the intervening 130 years—there have been remarkably few attempts to reaffirm or improve on the basic survey data concerning the primary orientations. Existing discussions typically depend upon measurements dating back at least 25 years and in part, either explicitly or implicitly, right back to Petrie's own. Nor have these data generally been considered in the broader context of the orientations of the numerous other structures—in particular, the smaller pyramids and mastabas, as well as the valley and mortuary temples—in the vicinity.

This paper presents the results of a week-long Total Station survey undertaken by the authors during December 2006 whose principal aim was to clarify the basic data concerning the orientation of each side of the three large pyramids and to determine, as accurately as possible, the orientations of as many as possible of the associated structures. The survey focused upon measurements of sequences of points along multiple straight and relatively well preserved structural segments, with best-fit techniques being used to provide the best estimate of their orientation, as opposed to simple triangulation between directly identified or extrapolated corners. This facilitates a richer interpretative approach taking account of possible variations in orientation targets and alignment techniques at different stages of construction, both of individual monuments and among the various monuments and monument complexes on the plateau.

We proceed to a comparative analysis between our data and those that have been published in the past, and to offer some preliminary interpretations, focusing on the following key questions, which motivated the survey:
1. Are the azimuths measured by Petrie and quoted ever since, often to a precision of 0.1 arc minutes, actually a fair reflection of the intended orientation of the main Giza pyramids?
2. What do the basic data suggest about which of the cardinal directions was used to determine the orientation?
3. What does the broader context of structure orientations on the Giza plateau suggest about the techniques and motives of orientation?

Our results do not directly challenge the idea that pairs of circumpolar stars were used to determine the primary orientation of the largest pyramids, but they do suggest that there is only a very slight difference in orientation (≈0.5′) between the north-south axes of Khufu's and Khafre's pyramids, that the sides of Khafre's are more perfectly perpendicular than those of Khufu's, and that the east-west axis is closer to true cardinality in both cases. The broader context of associated structures suggests that the east-west orientation in relation to sunrise or (in one case) sunset may have been a, or even the, key factor in many cases.

# 1 Introduction

Putative celestial alignments on the Giza plateau, embodied both in sightlines between the structures and in the orientations of individual structures, have been the subject of considerable debate over the preceding century, with a renewed burst of interest taking place during the past decade. This paper is primarily concerned with the orientations of individual structures rather than alignments between them (regarding which see Lehner 1983; Houdin 2006; Magli 2008, 2009a, 2009b).

The largest of the Giza pyramids, often referred to as the Great Pyramid, was built in the mid-26th century BC for the fourth-dynasty pharaoh Khufu, known to the Greeks as 'Cheops' It consists of approximately 2,300,000 blocks of stone, averaging more than two tonnes in weight (Lehner 1997: 108). While the pyramid of his son Khafre ('Khephren' to the Greeks) is slightly smaller than Khufu's and that of his grandson Menkaure ('Mycerinus') is only about a quarter the size (in terms of area), all three pyramids are imposing monuments.

Khafre was not Khufu's only son, and was not his immediate successor. Khafre's elder brother Djedefre ruled for a short period after the death of Khufu and Djedefre decided to build his pyramid north of the Giza Plateau at Abu Roash. Djedefre's pyramid is much smaller than the three Giza Plateau pyramids, had a steeper incline and is poorly preserved (Reisner 1942; Lehner 1997: 120–121); therefore its orientation will not be considered further in this paper.

Ancient Egyptian pyramids were usually built with subsidiary structures and therefore had a number of associated architectural elements. The concept of a pyramid complex was not finalized until early dynasty 4, from which time each pyramid complex consisted of six such elements: a temenos (sacred enclosure) wall, the main pyramid, one or more satellite pyramids, a mortuary temple, a valley temple at some distance, and a causeway that connected the mortuary and valley temples. A pharaoh's family members or favourite courtiers were frequently buried in mastabas (flat-roofed rectangular tombs) next to his complex. Each of the three large pyramids at Giza formed part of such a complex, with its causeway running down to a valley temple some 500m away to the east. Adjacent to Khafre's valley temple was the Sphinx temple and the Sphinx itself.

## 1.1 *Egyptian survey methods*

Within a broad pattern of orientation in relation to the River Nile, many Egyptian temples were aligned cardinally, especially in Old Kingdom times.[1] Yet it has long been recognized that three Giza pyramids were aligned to the cardinal points with remarkable accuracy (Petrie 1883; Dorner 1981; Haack 1984). On the question of how this was achieved, evidence other than that of the alignments themselves is fragmentary. No engineering documents or architectural plans have been found that give technical explanations as to how the ancient Egyptians aligned any of their temples or pyramids. No ancient Egyptian compasses have ever been discovered, nor has any other type of sophisticated survey equipment (Romer 2007: 318).

Some records do exist, however, that appear to provide clues about the process of surveying. Depictions of foundation ceremonies for sacred buildings appear on temple walls throughout ancient Egyptian history, the most famous of these being that of the so-called 'stretching of the cord' ceremony found, for example, in the Temple of Horus at Edfu dating to 237 BC (Fig 1a). Here the king and the goddess Seshat (who represents numeration, counting, and time-keeping) are each seen holding a post; a rope is attached to both posts. The accompanying text records that the king first determines where the posts are to be set by astronomical observation and then he and Seshat pull the rope tight. The stretching of the cord ceremony is also depicted more than two millennia earlier, in the 5th-dynasty Sun Temple of Niuserre at Abu Ghurab (Gallo 1998: 82; Belmonte 2001: S7) and, indeed, variations of this scene have been discovered in several temples and tombs from different periods, often accompanied by texts. These records clearly suggest that cords, posts, and astronomical observations were being used to align and measure buildings (Arnold 1991: 252–253; Belmonte 2001: S6–S10), and the fact that the stretching of the cord ceremony is mentioned on the 2nd-dynasty Palermo Stone (Belmonte 2001: S7, S9) makes it clear that such methods were already current at the time of the construction of the Giza pyramids.

---

[1] The recent work of Belmonte and his collaborators (Shaltout and Belmonte 2005; Belmonte and Shaltout 2006; Shaltout et al. 2007; Belmonte et al. 2008; Belmonte and Shaltout 2009) has revealed a much more elaborate picture in which the temples of the Nile valley and delta were oriented both topographically and astronomically, the astronomical alignments being of three types: cardinal or inter-cardinal (by sighting on northern stars and then rotating if necessary); solar/calendrical; and upon the two brightest stars, Sirius and Canopus.

Some of the texts describe how the king would gaze at the stars before determining the 'corners' of the temple. In order to determine the correct orientation he is assisted not only by Seshat but also by the god Thoth, who holds a *merkhet* (Fig 1b). This was 'an object used to recognize', or an 'indicator', and is generally assumed to have been a timepiece such as a *clepsydra* (water clock) or a sun/shadow clock (Zaba 1953: 16f). The numerous references to "gazing at the stars" have focused most attention on the clepsydra, but water clocks are inaccurate and, in any case, were purportedly invented by the sage Amenemhat a thousand years after the Giza pyramids were built (Cotterell *et al*. 1986; Isler 1991: 156). If the merkhet was a sun/shadow clock it was probably used as a gnomon (Isler 1991a: 157–158), with a cord and plumb bob suspended from it in order to determine the local vertical (Wells 1996: 37).

A *bay* (Fig. 1c) is a notched palm rib thought to have been used in conjunction with a merkhet. The inscription on a dynasty 26 merkhet and bay belonging to Hor, a 26th-dynasty 'Overseer of the Hour' priest, states that the merkhet "know(s) the motion of the two discs [i.e. sun, moon] and every star to its place" and that part of the bay's use was "for indicating the commencement of a feast, and for placing every person in his hour" (*ibid*.).[2] It has been suggested that the base of the bay was placed on the ground and the selected asterism was sighted through the sliced notch on the other end (*ibid*.). However, if the notched end of the bay were placed against the ground it could also be used to accurately specify the position of the sun's shadow, as depicted in Fig. 1d (Singh 1978: 90–91; Isler 1989: 198). Another of the texts referring to Thoth, which describes the construction of a palace dedicated to Re-Horakhty (the sun god Re-of-the-two-horizons), suggests more directly that sun/shadow measurements were made. It makes reference explicitly to the sun's disk and to the length and width of the shadow thrown by it (Zaba 1953: 16f).

From the various available accounts, then, it seems clear at the very least that both stellar and sun/shadow observations could form part of the process of orienting a palace or temple. But what strategies were used in practice?

## *1.2    Hypotheses as to how the orientations were determined*

Ever since Flinders Petrie carried out the first theodolite survey of the pyramids of the Giza Plateau in 1881 (Petrie 1883: 6), the precision with which their sides are oriented upon the true cardinal directions has fascinated scholars. What methods could have been used to determine the orientation? The possibilities depend upon which direction was used.

---

[2]    These particular instruments belonging to Hor are in the Egyptian Museum, Charlottenburg, Berlin, 14084/14085.

## 1.2.1 Sighting upon circumpolar stars in the north

Horizon observations are useless to the north and south because the celestial bodies scrape along the horizon rather than moving significantly up or down; however, in the northerly direction several methods are possible using devices to monitor the azimuths of circumpolar stars.

Petrie himself hypothesized, effectively, that people could have observed a star close to the celestial pole using a fixed vertical plumb line dropped from a height of some 15m. An observer, to the south, would track the east-west motion of the star by moving from side to side, keeping the star behind the plumb line, and marking the places on the ground from which it was seen to reach its elongations, 12 hours apart. The halfway position between of these two points would be due south of the plumb line (Petrie 1883: 211–212).[3] Several decades later, Edwards (1947: 209–211) suggested that a level 'observation wall' could have been constructed, with northern stars being measured rising and setting behind it, so that the 'mean direction' would be true north. A problem with Petrie's method is that the sky will not be dark for both observations, so that they would need to be made in different seasons. Edwards' stars, unless rising and setting a long way apart, will do so at very shallow angles, and again the rising and setting may not both take place in darkness. In any case, no remains of any putative stellar 'observation walls' have been discovered at any Egyptian temple or pyramid site.

Spence (2000: 320–324) proposed the novel hypothesis that the ancient Egyptians used two circumpolar stars to align the Great Pyramid: Kochab (β UMi) and Mizar (ζ UMa). She suggested that, during a formal evening ritual, the priests and the king used a giant merkhet to hang a plumb line and thereby determine when the lower star was directly beneath the upper star, this then being taken as the direction of true north. Spence theorized that, if this method had been used by the ancient Egyptians, the Great Pyramid would have been aligned in 2467 BC ± 5 years. This is approximately a century later than the date that would be accepted by most Egyptologists—see note 7. To support her hypothesis she undertook a comparative dating/alignment analysis on several other pyramids and temples that were built just prior to or after the Great Pyramid, on the assumption that the same pair of stars was used but that a slightly different azimuth was obtained owing to precession. The pyramids belonging to Snefru (dynasty 4) and Sahure and Neferirkare (dynasty 5) fell within the range of her scatter plot and fitted her hypothesis, but those built by Djoser (dynasty 3) and Unas (dynasty 5) did not. As with "observation walls", there is no evidence of the requisite measuring device. None of the merkhets discovered in Egypt has been of sufficient size.

Belmonte (2001) cast doubt on Spence's hypothesis, questioning why she had picked those particular stars, especially given that—being on opposite sides of the celestial pole—this would have meant that the hypothetical giant merkhet would have needed to accurately determine the vertical for a measurement subtending an angle of some 40°. Belmonte proposed instead that the stars Phecda (γ UMa) and Megrez (δ UMa) would have been a better choice because, both being in Ursa Major, they are not only much closer together but also formed part of a single ancient Egyptian constellation, *Meskhetiu* (the Foreleg of the Bull), which is mentioned in the Pyramid Texts dating to dynasty 5. This would have meant that the time when they stood one directly above the other in the sky could have been determined using an ancient Egyptian version of a measuring instrument known to have been used by Roman surveyors—a groma (Fig. 2a). Miranda *et al*. (2009) argue that its design might have been represented in Seshat's crown, as seen in foundation ceremony depictions (Fig. 2b), which bears a striking resemblance to a Roman groma. However, this idea is not supported either by literary evidence or by temple illustrations.

---

[3] "If a pile of masonry some 50 feet high was built up with a vertical side from North to South, a plumb-line could be hung from its top, and observations could be made, to find the places on the ground from which the pole-star was seen to transit behind the line at the elongations, twelve hours apart. The mean of these positions would be due South of the plumb-line, and about 100 feet distant from it; on this scale 15″ of angle would be about 1/10 inch, and therefore quite perceptible." (Petrie 1883: 211–212). Petrie's astronomical arguments are confused, and he takes no account of precession (ibid.: 126–127) but his method would apply to a star close to the celestial pole at the time.

A question left unanswered regarding the 'simultaneous transit' strategy for determining the north direction is on what grounds the ancient Egyptians made the choice of a particular pair of stars. For example, the choice of Kochab and Mizar would, as Spence has shown, have achieved the greatest prevision at the time of Khufu and Khafre, but it is inconceivable that these stars were chosen for this reason some 50 years earlier; instead, we would have to assume that the level of precision achieved by those pharaohs was fortuitous. If, on the other hand, the level of precision achieved at Giza was deliberate rather than fortuitous, then the surveyors must already have determined that a particular pair of stars, when aligned vertically in the sky, would provide a particularly good indication of true north. One conceivable strategy, paralleling the Edwards method using a single star, would be to observe the same pair of stars twice, first one above the other and then vice versa, extending the plumb-line down to the horizon to determine the azimuth in each case, and using the mean point between them. However, this not only runs into the same problem of having to make the two observations in different seasons, in order to get a dark sky in each case, but also (in the Spence case) of having to determine the vertical over an even wider arc down to the horizon; furthermore this is also a problem for one of the two observations in the Belmonte case—the one where both the stars are above the celestial pole. Belmonte (2012: 138) has more recently suggested that, since Phecda and Megrez were aligned upon the 'pole star' of the time—Thuban (α Dra)—they might well seemed natural choices for that reason. In that case, only their lower culmination would have offered the required precision.

### 1.2.2   Noon shadows

Neugebauer (1980) and Isler (2001: 163) have independently suggested that the pyramids could have been aligned by solar methods using a gnomon and measuring cord. Neugebauer proposed that measurements of the tip of the shadow of a pyramidal block could have been taken throughout a day, resulting in a curve to the north of the block (see Fig. 2c). If this did not intersect the extensions of the eastern and western sides at equal distances from the corners (C and D in the figure), then an adjustment to the orientation would be needed. Factors such as the levelness of the ground are critical.

### 1.2.3   Sunrise in the east

One of the authors (EN) has explored the possibility that the easterly direction could have been determined using sunrise observations. However, there are several problems with this, related to the precise determination of the astronomical equinox compounded with problems due to the altitude of the horizon (cf. Ruggles 1997). On the other hand, sun worship was becoming one of the most dominant aspects of Egyptian religion during dynasty 4. We shall discuss this further in what follows.

## *1.3   Nature and quality of the existing survey data*

Between 1880 and 1882, Petrie (1883) conducted an extraordinarily comprehensive, careful and detailed survey including the pyramids of Khufu, Khafre, and Menkaure, as well as two of the smaller Queens' pyramids of Khufu. Remarkably few independent surveys have been undertaken since that time. Those of J.H. Cole in the 1920s (Cole 1925) and Josef Dorner in the 1970s (Dorner 1981) were restricted to the pyramids of Khufu and Khafre.

Petrie and Cole surveyed using vernier theodolites, triangulating from different theodolite stations to obtain distance measurements (Petrie 1883: 10–21, 24; Cole 1925: 1–2)[4]. Dorner, working several decades later, aligned his theodolite along what he took to be the best casing lines, then took offset measurements along their length (Dorner 2005).

Recent years have brought great improvements in technology, most notably Electronic Distance Measurement (EDM) devices that are integral to modern Total Stations, and the Global Positioning System (GPS) network. In 1984 the Giza Plateau Mapping Project (GPMP) established its own triangulation grid over the entire plateau, and themselves undertook surveys of the Khufu and Khafre pyramids (Lehner and Goodman 1984; Lehner 1985a).

A concern with interpretations based upon azimuths quoted from the early surveys is that they pay too little attention to the subtleties of context. This is not to imply any criticism of the early archaeologists—indeed, Petrie's achievements in 1881 are truly remarkable—but relates simply to the limitations of the technology available to them.

One of the huge advantages of EDM devices is that they enable one efficiently to take measurements of a series of points at frequent intervals along a straight segment, for example of the edge of a casing layer or stone course, and then calculate a best-fit straight line to provide the best estimate of the azimuth. On a pyramid side, there may be many such segments, well preserved, at different levels. The early surveyors, on the other hand, would have found it impracticable (using just a theodolite and a tape or chain) to do more than measure the end points of such segments (for example, the corners of a pyramid to obtain a best estimate of the azimuth of the sides), triangulating them in from different survey stations in order to obtain a single best estimate of the azimuth of the line joining them.

As Petrie (1883: 37) was well aware, different casing and masonry levels may provide different insights into the intended orientation as well as the process of how this was adhered to during construction. Among the obvious questions are: how accurately did the foundations, the casing stones and then successive courses of stones in the pyramid itself accurately preserve the same orientation? Indeed, were the courses of stones in the pyramid straight and precisely perpendicular, and did successive courses retain the same orientation? If not, and if the precise alignment was so important, which structure in fact defined it? Or was the alignment constantly re-determined and possibly by differing means?

With the aim of clarifying the basic orientation data, the current authors undertook an intensive Total Station survey over a six-day period in December 2006. We collected orientation survey data from the pyramids of Khufu, Khafre, and Menkaure, five of the eight queens' and satellite pyramids associated with those pharaohs, the causeways of Khafre and Menkaure, Khafre's valley temple and the Sphinx temple, and a sample of monuments within the western, eastern, and southern cemeteries (mastaba fields) surrounding Khufu's pyramid. Alignment information for Menkaure's valley temple was also incorporated by courtesy of the Giza Plateau Mapping Project.[5]

In this paper we present the principal results of this survey and provide a comparative analysis between our data and those that have been published in the past. We proceed to review the existing interpretations in the light of these new data and conclude by presenting some tentative conclusions of our own.

---

[4] Cole (1925: 1–2) recounts that "Eight brass bolts were cemented into the rock round the base, one near each of the four corners, and, as these were not intervisible, four more were placed, one at about the middle of each side, in such a position that each point was visible from the adjacent points on each side of it. … The positions of these points were determined by means of precise traverse observations."

[5] These data, and also those relating to the Khafre workmen's village and the complex of Khentkawes (see below) have been supplied by GIS specialist Farrah Brown LaPan of the Giza Plateau Mapping Project.

## 2   The 2006 survey: methods and procedures

### *2.1   Survey methods*

The authors set out to measure as many as possible of the orientations of monumental structures on the Giza plateau during an intensive six-day period starting on 16 December 2006. The work was undertaken using a Leica TCR705 Total Station loaned from the School of Archaeology and Ancient History at the University of Leicester. The survey was conducted under the Supreme Council of Antiquities umbrella permission granted to the Giza Plateau Mapping Project (GPMP), a division of Ancient Egypt Research Associates (AERA).

Our broad strategy was to identify a set of survey stations over the plateau that would not only enable us to establish a robust triangulation network but also to optimize the accuracy of our sightings of reference points on structures around the site. A set of 11 survey stations was duly identified. The direction of true north was determined independently at each survey station by taking one or more sets of sun-azimuth readings (Ruggles 1999: 168–169) and a triangulation grid was established by direct measurements between intervisible stations together with sightings upon locatable control points on the GPMP triangulation grid set up in 1984 (Goodman and Lehner 2007; Goodman 2007). The latter was itself aligned to true north by astronomical means—Polaris observations (Goodman and Lehner 2007: 56; Goodman 2007: 98)—and we were able to verify that our own grid and the GPMP grid agree on the direction of true (astronomical) north to within our own margins of error (± 0.3′), thus eliminating the possibility that any significant systematic errors affected our entire survey.

The EDM reflector was taken to prearranged locations along the base of each pyramid or monument and the position of each point determined as accurately as possible from at least one of the survey stations. For the large pyramids we aimed to obtain up to 30 readings from each of the four sides (at distance intervals of approximately 3m to 10m), depending upon the availability of original casing or foundation bases, in order to obtain best estimates of structure orientations and deviations from straightness. For the remaining monuments, depending on their size, at least 3 readings (and preferably many more) were taken from as many sides as possible given their condition of preservation and the time and logistical constraints. Numerous repeated measurements and consistency checks were made in order to reduce random errors and eliminate gross errors. As a result, distance measurements are generally considered accurate to 0.01m and plate bearing measurements to 0.3′.

Best-fit straight lines through sets of points were determined using the least-squares method with perpendicular offsets (see http://mathworld.wolfram.com/LeastSquaresFittingPerpendicularOffsets.html). The accuracy of the deduced orientation of an alignment of measured points depends upon several factors, in particular:
- their distance from the survey station;
- the number of points from which a best fit can be determined; and
- the angle between the alignment and the line of sight from the survey station, in other words the degree to which the deduced spatial location of any point in the alignment is dependent upon the component of distance measurement and that of plate bearing angle.

In what follows our orientations are generally quoted to a precision of 0.1′, and generally considered accurate to within 1′.

The survey stations and the principal structures measured from them are indicated in Fig. 3.

## 2.2 Terminology and conventions

In this paper we are making a clear distinction between precision and accuracy, following Ruggles (1999: ix). Accuracy is used exclusively to refer to how well the modern measurement of an attribute (such as the orientation of a structure) conforms to its true value. "Precision" is used primarily to refer to the size of the units in which a measurement is quoted, as in a statement such as "this measurement is quoted to a precision of ten arc seconds but only appears accurate to about a minute". "False precision" occurs whenever the degree to which a measurement is quoted misrepresents the actual accuracy of the reading. The term "precision" is also used more loosely to refer to the degree to which a structure was aligned upon an intended target, and to which the present-day orientation reflects the original one.

We also take "altitude" exclusively to mean the vertical angle between a viewed point and the horizontal plane through the observer (as opposed to elevation, which is taken to mean the height of a location above sea level). Finally, "azimuth" means "true azimuth" throughout.

When referring to the orientation of one side of a square or rectangular structure, or to the mean orientation of the four sides of such a structure, we express this in terms of a **deviation from cardinality (DFC)** expressed as a positive angle if the deviation is clockwise (CW) or a negative angle if the deviation is counterclockwise (CCW). Thus for example:

| **Mean deviation from cardinality (DFC)** | **Structure** | **N azimuth** | **E azimuth** | **S azimuth** | **W azimuth** |
|---|---|---|---|---|---|
| –3° | 3° CCW | 357° | 87° | 177° | 267° |
| –0° 3.5′ | 0° 3.5′ CCW | 359° 56.5′ | 89° 56.5′ | 179° 56.5′ | 269° 56.5′ |
| +0° 0.0′ | Cardinal | 0° 0.0′ | 90° 0.0′ | 180° 0.0′ | 270° 0.0′ |
| +0.5° | 0° 3.5′ CW | 0° 3.5′ | 90° 3.5′ | 180° 3.5′ | 270° 3.5′ |
| +3° | 3° CW | 3° | 93° | 183° | 273° |

## 2.3 Structures surveyed

### 2.3.1 Khufu's pyramid and its associated structures

Khufu's Valley Temple and much of its causeway have been destroyed but three smaller pyramids possibly built for Khufu's queens survive immediately to the east of the Great Pyramid. There was also an even smaller satellite pyramid in this vicinity, but only parts of its foundations survive. Adjacent to Khufu's pyramid complex on three sides are arrays of mastabas (Reisner 1942: 9–13): the western cemetery or western mastaba field (WC/WMF), the eastern cemetery or eastern mastaba field (EC/EMF), and the southern cemetery (SC), arranged in an east-west line.

*Queens' pyramids*

We were only able to collect reliable data on the central one of the three queens' pyramids, to which we refer as KQ2, owing mainly to the eroded bases of the other two. (This was despite the fact that visibility considerations constrained us to place one of our survey stations atop the southernmost pyramid KQ3.)

*Mortuary temple*

We were only able to collect reliable data on the southern part of the eastern side, opposite the northernmost queen's pyramid KQ1.

*Western cemetery*

The WC/WMF was excavated in the early twentieth century by Hermann Junker, George Reisner, Clarence Stanley Fisher, and Ernesto Schiaparelli. The oldest part was built in the early part of Khufu's reign, and includes the tomb of Khufu's chief architect, Hemiunu (Reisner 1942: 9–12, no. G 4000), as well as some of Khufu's important sons (ibid.: G 1200, G 2000, G 2100). For a plan see Reisner (1942: map 2); and also the Giza Archives website of the Museum of Fine Arts, Boston (http://www.gizapyramids.org/). The other three main concentrations of tombs in the WC/WMF were constructed later in Khufu's reign and throughout dynasties 5 and 6 (Reisner 1942: 13). We collected orientation data from a small sample of the mastabas that were constructed in the early part of Khufu's reign, situated immediately to the east of the Tomb of Hemiunu.

*Eastern cemetery*

The EC/EMF is located to the east of KQ1–KQ3 and includes tombs for Khufu's daughters and wives. The majority of the mastabas in the EC/EMF were built in the middle to later part of Khufu's reign, meaning they were begun well after the earlier mastabas in the WC/WMF. Additional burials were added to the EC/EMF throughout the rest of the Old Kingdom (c. 2686 to 2160 BC).[6] Almost 1500 years later, in the Saite, Ptolemaic, and Roman periods, even more burials were added (Reisner 1942: 13). We collected data from sixteen mastabas located in four N–S rows immediately to the east of the queens' pyramids.

*Southern cemetery*

The last group comprises a single row of nine mastabas built for nobles and situated to the south of Khufu's pyramid (Fig. 5b). Although adjacent to the Great Pyramid, these mastabas were not built during Khufu's domination, but rather during the reigns of Khufu's son and grandson, the pharaohs Khafre and Menkaure (Reisner 1942: 12–20). They were poorly preserved and bear much reconstruction, so data collected from this group may not be as precise as the other structures in Khufu's complex. We collected data from the westernmost three of these mastabas.

### 2.3.2   Khafre's pyramid and its associated structures

Khafre's pyramid has a small satellite pyramid to its south, and a causeway leading ESE down to his valley temple and the Sphinx temple adjacent to it.

*Satellite pyramid*

This small pyramid is poorly preserved and its foundation almost completely destroyed. We were only able to collect reliable survey data from one side.

---

[6] Here and elsewhere we quote dates from Malek (2000) and Shaw (2000: 479–483) for reference, while fully acknowledging that debate continues among Egyptologists concerning absolute chronologies. See, for example, Belmonte and Shaltout (2009): appendix I. These dates are not critical to any of the arguments given in this paper.

*Causeway*

The central part of the causeway is sufficiently well preserved for a good estimate to be made of the intended orientation. The southern edge is less well defined than the northern edge.

*Valley temple and Sphinx temple*

Khafre's valley temple (Fig. 4) is located at the eastern of the causeway and was split into lower and upper sections. The Sphinx temple (the Sphinx itself is to its west) abuts the northern side of the valley temple; the two temples share a similar alignment and give the impression of being one building split into two parts by a passageway. The interior plan of the Sphinx temple is unique in that it includes both an eastern and western sanctuary along the central axis of the temple.

### 2.3.3   Menkaure's pyramid and its associated structures

Like Khufu's pyramid, Menkaure's has three smaller queens' pyramids, but they are on its south side. The causeway runs east.

*Queens' pyramids*

These three pyramids, to which we refer as MQ1–MQ3 running from west to east, are badly deteriorated and we were only able to collect data from all four sides in the case of MQ1.

*Causeway*

The outer edge of the pavement on the northern side of the western half of the causeway is sufficiently well preserved for a good estimate to be made of the intended orientation.

### 2.3.4   Summary

In summary, then, the structures targeted for our survey were as follows:

- The pyramid of Khufu (Great Pyramid): each of the four sides.
- The central one of the three pyramids of Khufu's queens.
- Khufu's mortuary temple.
- Khufu's western cemetery (Western Mastaba Field): a sample of early mastabas immediately to the east of the Tomb of Hemiunu.
- Khufu's eastern cemetery (Eastern Mastaba Field), a selection of the early mastabas closest to Khufu's pyramid.
- Southern cemetery (mastabas believed to have been mostly constructed by courtiers of Khafre and Menkaure, not Khufu): the westernmost three in the row of nine mastabas.

- The pyramid of Khafre: each of the four sides.
- Khafre's satellite pyramid.
- Khafre's causeway.
- The Sphinx temple.
- Khafre's valley temple.

- The pyramid of Menkaure: each of the four sides.
- The three pyramids of Mankaure's queens. We refer to these as MQ1–MQ3, running from west to east.
- Menkaure's causeway.

# 3    The 2006 survey: basic results

In this section we present the basic survey results for each of the oriented structures just listed. For the main pyramids we treat each of the four sides independently, but even in the case of a single side there are various structures that might be taken as providing an indication of the intended orientation. The main possibilities are:

- The *casing foundation*: a series of stone blocks, extending outward around the pyramids, which supported the casing stones. Readings were taken on the outer edge, specified as being along the base or top of this edge where the distinction was deemed significant. Petrie refers to the casing foundation as the 'casing in situ'.

- The *outer foundation*: a lower level of stone blocks supporting the casing foundation, which extends yet further out from the pyramid. Readings were taken on the outer edge. Petrie refers to the outer foundation as the 'pavement'.

- The stone *courses* of the pyramid, numbered from the 1st upwards. Readings were taken on the outer edge (top), or along the bottom of a stone course, as circumstances dictated. In one case at Menkaure's pyramid, the best indication of the intended orientation was deemed to be a straight line etched along the stones as part of the construction process.

In each case, we sought to identify straight segments of the edge of the casing foundation, outer foundation or stone courses that were sufficiently undamaged to yield a reasonable estimate of the original orientation. Points along the segment were then measured at intervals of between 3m and 10m. Khafre's pyramid is surrounded by alignments of postholes, but we did not survey these because this work had already been undertaken by Lehner (1983; 1985a; 1986); similar postholes surrounding Khufu's pyramid are in such a poor state of preservation that they seem to be randomly scattered.

The results are summarised in the six tables that follow (Tables 1–6). The following notes apply to all of these tables.

1. Without loss of generality, azimuths are expressed in one of the two possible directions—eastwards for east-west alignments and northwards for north-south alignments.

2. Azimuths and deviations from cardinality (DFCs) are only quoted to a precision of 0.1´ where they are based upon a best-fit of some 9 or more points. Otherwise they are quoted to a precision of 1´. Where the azimuths is based upon only 3 points, the entry is given in light type to indicate that it may be optimistic to expect even this level of accuracy.

3. Italicised rows indicate segments that are subsequently considered together.

4. Bolded rows indicate data considered to be of particular interest or importance.

## 3.1 The Great Pyramid of Khufu

The basic data for the Great Pyramid are shown in Table 1a. Included in these data are the surviving three of four brass corner markers put in place by earlier surveyors.

**Table 1a: The Great Pyramid of Khufu: basic data**

| Side | Indication | Part | No. of points | Best-fit azimuth | Deviation from cardinality (DFC) | Notes |
|---|---|---|---|---|---|---|
| North | Outer foundation | | 7 | 89° 54´ | –0° 06´ | a |
| **North** | **Extended casing below entrance, front** | | **9** | **89° 56.4´** | **–0° 03.6´** | |
| North | Brass corner markers | | 2 | 89° 56.4´ | –0° 03.6´ | c |
| North | Casing foundation, base | Western | 6 | 89° 47´ | –0° 13´ | b |
| North | Reconstructed casing foundation | Western –Central | 5 | 89° 56´ | –0° 04´ | |
| *East* | *Casing foundation* | *Northern* | *3* | *0° 02´* | *+0° 02´* | |
| East | Casing foundation | Central | 10 | 359° 58.1´ | –0° 01.9´ | |
| East | Casing foundation | Southern | 4 | 359° 52´ | –0° 08´ | |
| **East** | **Casing foundation** | | **17** | **359° 56.6´** | **–0° 03.4´** | |
| South | Casing foundation | Eastern | 10 | 89° 58.4´ | –0° 01.6´ | |
| South | Casing foundation | Western –Central | 11 | 90° 00.5´ | +0° 00.5´ | |
| **South** | **Casing foundation** | | **21** | **89° 59.5´** | **–0° 00.5´** | |
| West | Outer foundation | Southern | 5 | 359° 53´ | –0° 07´ | |
| West | *Casing foundation, base* | *Southern* | 8 | 359° 54´ | –0° 06´ | |
| West | *Casing foundation, base* | *Northern* | 7 | 359° 58´ | –0° 02´ | |
| **West** | **Casing foundation, base** | | **15** | **359° 56.3´** | **–0° 03.7´** | |
| West | Casing foundation, top | Central | 7 | 359° 54.6´ | –0° 05.4´ | |
| West | Brass corner markers | | 2 | 359° 52.7´ | –0° 07.3´ | c |

**Notes**
a  The identification of this as the outer foundation is not certain.
b  The identification of this as the casing foundation base is questionable.
c  In the case of the brass corner markers, four independent readings were taken of each pairwise alignment, consistent to within 0.2´ in each case, so the quoted precision is considered justified.

Data from the west and south sides indicate that the edge of the casing foundation base on each of these sides was slightly convex. On the west side, the difference in azimuth at the two ends is about 4´, and the data from the south side are consistent with this figure. The data from the north side suggest a greater degree of convexity, but see (2) below. This is in contrast to the well-documented concavity of the stone faces of the pyramid itself (Edwards 1947: 222; Maragioglio and Rinaldi 1965: 16, 104; Isler 1983); only the casing foundation on the remaining, east, side seems to have been concave.

Comparing the orientations of the edge of the casing foundation from the four sides reveals that:

1. The edges of the casing foundation on the east and west sides are almost exactly parallel, the mean DFCs being –0° 03.4´ and –0° 03.7´ respectively.
2. The questionable segment of original casing foundation base on the north side yields an anomalous DFC value of –0° 13´, suggesting that it was not in fact genuine. On the other hand, the extended casing below the entrance yields –0° 03.6´, as does (within greater margins of error) a reconstructed segment of the casing foundation edge.
3. The orientation of the casing foundation edge on the south side is significantly different, yielding a mean DFC of just –0° 00.5´.
4. (2) and (3) together show that the mean orientations of the casing foundation edges on the north and south sides differ from each other by at least 3´, converging towards the west.

We conclude that the intended orientations of the four sides of the Great Pyramid were as follows:

**Table 1b: The Great Pyramid of Khufu: summary data**

| Wall | Mean deviation from cardinality (DFC) |
|:---:|:---:|
| North | –0° 03.6′ |
| East | –0° 03.4′ |
| South | –0° 00.5′ |
| West | –0° 03.7′ |

Regarding the brass corner markers, the orientation obtained from those on the north side exactly matches that obtained from the extended casing below the entrance. In contrast, those on the west side yield an orientation (–0° 07.3´) markedly different from the mean orientation of the casing foundation base (–0° 03.7´). On the other hand, both corner-marker azimuths are within 0.3´ of the core masonry azimuths quoted by Petrie (1883: 38), namely –03.3´ for the north side and –07.5´ for the west side. There is some confusion in the literature about the origins of these markers,[7] but this seems to confirm that they coincide with Petrie's (1883: 37) rock-cut sockets.

---

[7] The brass survey plugs (along with the iron post on top of the Great Pyramid) were set in place by the astronomer David Gill during his return from Mauritius to observe the Transit of Venus of 1874 (Romer 2007: 41; Wilma Wetterstrom, priv. comm... 2012; note that Lehner (1985a: 27) states erroneously that they were set in place by Petrie). Petrie, who avoided using metal markers because they were likely to be stolen or damaged (1883: 20), found one of Gill's plugs missing and replaced it with a drilled hole filled with plaster. Cole (1925: 1–2) cemented brass bolts into the rock, "one near each of the four corners", in positions determined independently. Maragioglio and Rinaldi (1965: 14–15) confirm finding bronze nails fixed in the NE and NW corner-sockets, "while in the south-west corner-socket there remains only the hole in which the nail was fixed and in the south-east one nothing remains to show the place of the nail" (ibid.: 14). When Lehner and Goodman cleared the SW socket they located a hole, 4 cm square, where the brass survey plug would have been located (Lehner 1985a: 27), and replaced the plug in the SW corner (Wetterstrom, priv. comm., 2012).

## *3.2  Structures associated with Khufu's pyramid*

The basic data from the pyramid complex of Khufu are presented in Table 2a.

For ease of interpretation, we refer to tombs in the western cemetery as WC-1a etc, with numbers increasing east to west and letters increasing north to south, so that (e.g.) WC–3b is the second tomb, counting from the north end, in the third N–S row, counting from the west. We include cross-references to Reisner's numbering scheme (1942: Map 2). For example, WC–3a is Reisner's G 4360.

Likewise, we refer to the measured tombs in the eastern cemetery (Fig. 5a) as EC-1a etc, with numbers increasing east to west and letters increasing north to south, so that (e.g.) EC–3b is the second tomb, counting from the north end, in the third N–S row, counting from the west (closest to the Great Pyramid). Again, we include cross-references to Reisner's numbering scheme (1942: Map 3); for example EC–1a is G 7110. In this area, mastabas in each row are paired, so that "a" and "b" in each row, and again "c" and "d" in each row, are enclosed within a single rectangular wall. Where we have collected the relevant data, we give the best-fit west wall orientations of such 'double' structures, e.g. EC-1c/1d.

We refer to the southern mastabas (Fig. 5b) as SC-1, SC-2, etc running from west to east, matching Junker's (1955) terminology M.I, M.II etc. (see also The Giza Archives of the Museum of Fine Arts, Boston (http://www.gizapyramids.org/) and Reisner's GIS, GIIS, GIIIS, … GIXS (Reisner 1942: 17).

**Table 2a: Great Pyramid of Khufu, associated structures: basic data**

| Structure | Side | Indication | No. of points | Best-fit azimuth | Deviation from cardinality (DFC) | Notes |
|---|---|---|---|---|---|---|
| Mortuary temple | East | (Southern part) | 4 | 359° 59´ | –0° 01´ | |
| Queen's pyramid KQ2 | East | Casing foundation, top | 4 | 359° 45´ | –0° 15´ | |
| *Queen's pyramid KQ2* | *West* | *Casing foundation, base* | *3* | *0° 26´* | *+0° 26´* | |
| *Queen's pyramid KQ2* | *West* | *Casing foundation, top* | *3* | *0° 50´* | *+0° 50´* | |
| *Mastaba EC-1a (G 7110)* | *West* | *Casing foundation, base* | *3* | *359° 47´* | *–0° 13´* | *LS* |
| *Mastaba EC-1b (G 7120)* | *West* | *Casing foundation, base* | *3* | *0° 04´* | *+0° 04´* | *LS* |
| **Mastaba EC-1a/1b (G 7110/7120)** | **West** | **Casing foundation, base** | **6** | **359° 42´** | **–0° 18´** | **LS** |
| Mastaba EC-1b (G 7120) | South | Casing foundation, base | 3 | 89° 19´ | –0° 41´ | |
| Mastaba EC-1c (G 7130) | South | Casing foundation, base | 3 | 90° 12´ | +0° 12´ | |
| Mastaba EC-1d (G 7140) | North | Casing foundation, base | 3 | 88° 58´ | –1° 02´ | |
| *Mastaba EC-1c (G 7130)* | *West* | *Casing foundation, base* | *3* | *359° 22´* | *–0° 38´* | *LS* |
| *Mastaba EC-1d (G 7140)* | *West* | *Casing foundation, top* | *4* | *359° 50´* | *–0° 10´* | *LS* |
| **Mastaba EC-1c/1d (G 7130/7140)** | **West** | **Casing foundation, base** | **7** | **359° 27´** | **–0° 33´** | **LS** |
| Mastaba EC-2a/2b (G 7210/7220) | West | 2nd course | 5 | 359° 33´ | –0° 27´ | LS |
| *Mastaba EC-2b (G 7220)* | *South* | *Casing foundation, base* | *3* | *90° 19´* | *+0° 19´* | |

**Table 2a (ctd)**

| Structure | Side | Indication | No. of points | Best-fit azimuth | Deviation from cardinality (DFC) | Notes |
|---|---|---|---|---|---|---|
| Mastaba EC-3b (G 7320) | South | Casing foundation, base | 3 | 90° 17´ | +0° 17´ | |
| *Mastaba EC-3c (G 7330)* | *West* | *Casing foundation, base* | *3* | *359° 55´* | *–0° 05´* | *LS* |
| *Mastaba EC-3d (G 7340)* | *West* | *Casing foundation, base* | *3* | *0° 41´* | *+0° 41´* | *LS* |
| **Mastaba EC-3c/3d (G 7330/7340)** | **West** | **Casing foundation, base** | **6** | **0° 18´** | **+0° 18´** | **LS** |
| Mastaba EC-3e (G 7350) | North | North wall | 3 | 87° 19´ | –2° 41´ | |
| Mastaba EC-4b (G 7420) | South | Casing foundation, base | 3 | 90° 6´ | +0° 6´ | |
| Mastaba EC-4b (G 7420) | West | Casing foundation, base | 3 | 0° 30´ | +0° 30´ | LS |
| Mastaba EC-4c (G 7430) | North | Casing foundation, base | 3 | 90° 24´ | +0° 24´ | |
| Mastaba EC-4c/4d (G 7430/7440) | West | Casing foundation, base | 4 | 0° 57´ | +0° 57´ | LS |
| Mastaba EC-4e (G 7450) | North | Casing foundation, base | 3 | 88° 29´ | –1° 31´ | |
| | | | | | | |
| Mastaba SC-1 | North | Casing foundation, base | 4 | 90° 43´ | +0° 43´ | |
| Mastaba SC-1 | East | Casing foundation, base | 5 | 359° 50´ | –0° 50´ | LS |
| Mastaba SC-1 | South | Casing foundation, base | 4 | 87° 43´ | –2° 17´ | |
| Mastaba SC-1 | West | Casing foundation, base | 5 | 359° 43´ | –0° 17´ | LS |
| Mastaba SC-3 | North | Casing foundation, base | 3 | 89° 57´ | –0° 03´ | |
| Mastaba SC-3 | East | Casing foundation, base | 4 | 359° 08´ | –0° 52´ | LS |
| Mastaba SC-3 | South | Casing foundation, base | 4 | 88° 11´ | –1° 49´ | |
| Mastaba SC-3 | West | Casing foundation, base | 5 | 0° 02´ | +0° 02´ | LS |
| Mastaba SC-4 | North | Casing foundation, base | 4 | 90° 21´ | +0° 21´ | |
| Mastaba SC-4 | East | Casing foundation, base | 7 | 359° 48´ | –0° 12´ | LS |
| Mastaba SC-4 | South | Casing foundation, base | 4 | 90° 07´ | +0° 07´ | |
| Mastaba SC-4 | West | Casing foundation, base | 5 | 358° 56´ | –1° 04´ | LS |
| | | | | | | |
| Mastabas WC–2b/3b (G 4250/4350) | North | Casing foundation, base | 3 | 88° 14´ | –1° 46´ | |
| Mastaba WC-2e (G 4220) | West | Casing foundation, base | 3 | 356° 52´ | –3° 08´ | LS |
| Mastaba WC-3a (G 4360) | North | Casing foundation, base | 3 | 87° 01´ | –2° 59´ | |
| Mastaba WC-3a (G 4360) | West | Casing foundation, base | 3 | 357° 32´ | –2° 28´ | LS |
| Mastaba WC-4a (G 4460) | North | Casing foundation, base | 3 | 86° 49´ | –3° 11´ | |
| Mastaba WC–4a (G 4460) | West | Casing foundation, base | 3 | 358° 31´ | –1° 29´ | LS |
| Mastaba WC-4b (G 4450) | West | Casing foundation, base | 3 | 357° 44´ | –2° 16´ | LS |

**Notes**
LS    Longer side (of mastabas)

These data reflect the poor state of preservation of the structures involved, and in many cases we were only able to obtain lines of 3 points on any given side, with consequent uncertainties in the azimuth values obtained. In the case of the western cemetery, particularly, we obtain values for the DFC varying from –1.5° to –3.2° on different mastaba sides, with no reduction in this range if we restrict ourselves to the longer (west and east) sides. Yet, given that our data set comprises seven independent values obtained from six different tombs, it is safe to conclude that the DFC, averaging out at around –2.4°, was not only significantly but also deliberately different from that of the Great Pyramid itself.

If we exclude the remaining azimuth estimates that are based on only 3 values, then a more coherent picture begins to emerge in relation to the remaining structures in the Khufu complex. The data are summarised in Table 2b. Values from the shorter (north and south) sides of mastabas in the southern cemetery are included where lines of more than 3 points have been measured, but are placed in brackets to reflect the fact that they may be of secondary significance.

**Table 2b: Great Pyramid of Khufu and associated structures: summary data**

| Structure | Approx. foundation date * | Deviation from cardinality (DFC) | | | | | |
|---|---|---|---|---|---|---|---|
| | | North side | East side | South side | West side | Mean (all sides) | Mean (E & W sides) |
| **Great pyramid** | **2587 BC** | –0° 04′ | –0° 03′ | –0° 01′ | –0° 04′ | –0° 03′ | –0° 03′ |
| Khufu mortuary temple | 2587 BC | | 0.0° | | | | |
| Queen's pyramid KQ2 | 2585 BC | | –0.3° | | | | |
| EC row 1 (7100) | 2583 BC | | | | –0.5° to –0.3° | | |
| EC row 2 (7200) | 2583 BC | | | | –0.5° | | |
| EC row 3 (7300) | 2583 BC | | | | +0.3° | | |
| EC row 4 (7400) | 2583 BC | | | | +1.0° | | |
| SC–1 | 2535 BC | (+0.7°) | –0.8° | (–2.3°) | –0.3° | –0.7° | –0.5° |
| SC–3 | 2535 BC | | –0.9° | (–1.8°) | 0.0° | | –0.5° |
| SC–4 | 2535 BC | (+0.4°) | –0.2° | (+0.1°) | –1.1° | –0.2° | –0.7° |

**Notes**
* See note 7. The dates quoted here (purely for reference and not critical to the arguments presented) are arrived at by taking Shaw's (2000: 480) most likely date for Khufu's accession, 2589 BC, and then tentatively assuming that the main pyramid, mortuary temple and western cemetery were started a couple of years into the reign while the queens' pyramids and eastern cemetery were started a little later. Some SC tombs were built during Khafre's reign (2558 – 2532 BC according to Shaw) while others were built during Menkaure's reign (2532 – 2503 BC), so we have adopted an approximate starting date for these of 2535 BC.

---

The data from the western part of the southern cemetery indicate that the mastabas here were oriented consistently, with a mean DFC of around –0.5°, although the eastern and western sides of individual mastabas were off-parallel by up to c. 1.0°.

From the remaining data we can tentatively conclude the following:
1. The orientation of the mortuary temple reflected that of the Great Pyramid itself.
2. The orientation of KQ2 was significantly different, with a DFC of some –0.3°.

3. The orientations of the westernmost two rows of mastabas in the eastern cemetery, closest to the Khufu's pyramid and his queens' pyramids, follow a similar orientation, with DFCs between –0.3° and –0.5°.
4. The orientations of the next two rows of mastabas, in contrast, deviated from cardinality in the opposite sense, with the DFC of row 4 (+1.0°) being significantly greater than that of row 3 (+0.3°).

On the basis of these data we can draw two general conclusions.

First, none of the burial structures of Khufu's sons and advisors achieved the degree of perfection to which his own pyramid was cardinally oriented. The orientation of KQ2 suggests that the same may have been true of those of his three wives.

Second, there appears to have been a systematic variation from west to east, with the mastabas of the western cemetery having DFCs between around –3.0° and –1.5°, those of the southern cemetery around –0.5°, and those of the eastern cemetery varying from around –0.5° on the western side to around +1.0° further to the east.

## 3.3 The Pyramid of Khafre

The basic data for Khafre's Pyramid are shown in Table 3a.

**Table 3a: The pyramid of Khafre: basic data**

| Side | Indication | Part | No. of points | Best-fit azimuth | Deviation from cardinality (DFC) | Notes |
|---|---|---|---|---|---|---|
| *North* | *Outer foundation* | *Western* | *3* | *89° 01´* | *–0° 59´* | *T1* |
| **North** | **Casing foundation** | | 13 | **89° 56.2´** | **–0° 03.8´** | **T1** |
| *North* | *1st course, outer edge* | *Western* | *7* | *90° 01´* | *+0° 01´* | *T1* |
| *North* | *1st course, outer edge* | *Central* | *6* | *89° 33´* | *–0° 27´* | *T1* |
| *North* | *1st course, outer edge* | *Eastern* | *3* | *89° 41´* | *–0° 19´* | *T1* |
| **North** | **1st course, outer edge** | | 16 | **89° 58.7´** | **–0° 01.3´** | **T1** |
| **East** | **Casing foundation, base** | | 19 | **359° 56.0´** | **–0° 04.0´** | **T4** |
| *East* | *1st course, outer edge* | *Northern–Central* | *14* | *359° 55.9´* | *–0° 04.1´* | *T4* |
| *East* | *1st course, outer edge* | *Central–Southern* | *4* | *359° 53´* | *–0° 07´* | *T4* |
| *East* | *1st course, outer edge* | *Southern* | *6* | *0° 19´* | *+0° 19´* | *T4* |
| **East** | **1st course, outer edge** | | 24 | **359° 58.8´** | **–0° 01.2´** | **T4** |
| South | Outer foundation | Western | 11 | 89° 57.2´ | –0° 02.8´ | T4 |
| **South** | **Casing foundation, base** | | 11 | **89° 54.2´** | **–0° 05.8´** | **T4** |
| South | Casing foundation, base | Western | 6 | 89° 56´ | –0° 04´ | T5 |
| **South** | **1st course, outer edge** | **Western** | **6** | **89° 35´** | **–0° 25´** | **T4** |
| South | 2nd course, bottom | Eastern | 5 | 90° 40´ | +0° 40´ | T4 |
| *South* | *2nd course, bottom* | *Central–Eastern* | *3* | *90° 08´* | *+0° 08´* | *T4* |
| South | 2nd course, bottom | Central | 5 | 90° 07´ | +0° 07´ | T4 |

**Table 3a (ctd)**

| Side | Indication | Part | No. of points | Best-fit azimuth | Deviation from cardinality (DFC) | Notes |
|---|---|---|---|---|---|---|
| West | Escarpment wall, bottom | | 17 | 359° 45.4´ | –0° 14.6´ | T1 |
| West | Outer foundation | Southern | 7 | 359° 55´ | –0° 05´ | T5 |
| West | Outer foundation | Southern | 9 | 359° 54´ | –0° 06´ | T1 |
| West | Outer foundation | Northern | 5 | 359° 50´ | –0° 10´ | T1 |
| West | *Casing foundation* | *Southern* | *4* | *359° 57´* | *–0° 03´* | *T5* |
| West | *Casing foundation* | *Northern* | *7* | *359° 55´* | *–0° 05´* | *T5* |
| **West** | **Casing foundation** | | **11** | **359° 55.8´** | **–0° 04.2´** | T5 |
| West | *1st course, outer edge* | *Southern* | *9* | *359° 54´* | *–0° 06´* | *T1* |
| West | *1st course, outer edge* | *Northern* | *8* | *0° 00´* | *0° 00´* | *T1* |
| **West** | **1st course, outer edge** | | **17** | **359° 55.0´** | **–0° 05.0´** | T1 |
| West | *1st course, outer edge* | *Southern* | *6* | *359° 49´* | *–0° 11´* | *T5* |
| West | *1st course, outer edge* | *Northern* | *5* | *359° 59´* | *–0° 01´* | *T5* |
| **West** | **1st course, outer edge** | | **11** | **359° 55.2´** | **–0° 04.8´** | T5 |

**Notes**
T1 Measurements taken from survey station 1 (see Fig. 3).
T4 Measurements taken from survey station 4 (see Fig. 3).
T5 Measurements taken from survey station 5 (see Fig. 3).

---

At the Khafre pyramid we have data from both the casing foundation and the first course on all four sides.

The casing foundation data reveal that:
1. As at Khufu's pyramid, the edges of the casing foundation on the east and west sides are almost exactly parallel, the DFCs here being –0° 04.0´ and –0° 04.2´ respectively.
2. As seems to be the case at Khufu's pyramid, the edge of the casing foundation on the north side is almost exactly perpendicular to that on the east and west sides, yielding a DFC of –0° 03.8´.
3. Again as in the case of Khufu's pyramid (but not so pronounced), the orientation of the casing foundation edge on the south side is somewhat different, yielding a mean DFC of –0° 05.8´.
4. Unlike at Khufu's pyramid, the slightly different alignments of the casing foundation edges on the north and south sides converge towards the east.

The data from the first course show additionally that:
5. On the west side, the first course of stones is aligned with the casing foundation base to a precision of better than 1.0´. On the east side, the northern and central part of the first course is so closely aligned with the casing foundation base that the two orientations are indistinguishable, but towards the southern end the first course orientation deviates considerably. On the north side the difference between the two levels is some 2.5´, with the mean orientation of the stone course actually being aligned more closely to the true cardinal direction. However on the south side the lower courses of stones seem to have less closely aligned with the casing foundation base.
6  On the south side there is clear evidence that the lower stone courses were concave, but on the west side the first course was slightly convex. The north side also seems to have been slightly convex, but with a possible deviation towards the east end. The evidence from the east side is less clear, given a marked deviation in orientation towards the southern end. In sum, there is no clear evidence that the lower faces of the Khafre pyramid were generally concave, and indeed Maragioglio and Rinaldi (1966) do not remark upon this, in contrast to the other two main pyramids.

The data from the outer foundation show additionally that:
7. On the south and west sides, where sufficient readings have been taken, the outer foundation orientation is within 5′ of the casing foundation orientation. On the south side it is actually closer to true cardinality, with a DFC of just –2.8′.

Additional data from the second course on the southern side indicate that this was skewed significantly clockwise, with a positive DFC perhaps as great as c. +40′ at the eastern end.

Various data show significant departures from straightness on all sides and at various levels, particularly in the stone courses. The outer foundation and casing foundation appear to have been somewhat concave on the west side at least.

We note finally that at the Khafre pyramid we were able to survey a line of more than 10 points along the same structure (west side, outer edge of the first course) twice, independently, from different survey stations. This confirmed the accuracy of our results to within 0.2′ for lines of 10 or more points. Two similarly repeated shorter lines of points also confirm the accuracy of our results within the expected margins.

To summarise, taking the casing foundation data as the best available indication of the intended orientation, as at Khufu's pyramid, we obtain the results shown in column 2 of Table 3b. We include for comparison the azimuths obtained from the first stone courses.

**Table 3b: The Pyramid of Khafre: summary data**

| Wall | Mean deviation from cardinality (DFC) | |
|---|---|---|
| | **As determined from casing foundation base** | *Comparative data from 1st course, outer edge* |
| North | –0° 03.8′ | –0° 01.3′ |
| East | –0° 04.0′ | –0° 15.4′ |
| South | –0° 05.8′ | –0° 25′ |
| West | –0° 04.2′ | –0° 04.9′ |

## 3.4   Structures associated with Khafre's pyramid

The basic data from the pyramid complex of Khafre are presented in Table 4a.

**Table 4a: Pyramid of Khafre, associated structures: basic data**

| Structure | Side | Indication | Part | No. of points | Best-fit azimuth | Deviation from cardinality (DFC) | Notes |
|---|---|---|---|---|---|---|---|
| Satellite pyramid | North | Outer wall | | 3 | 90° 04′ | +0° 04′ | |
| Causeway | North | North edge | | 8 | 103° 26′ | +13° 26′ | |
| Causeway | South | South edge | | 4 | 103° 33′ | +13° 33′ | |

**Table 4a (ctd)**

| Structure | Side | Indication | Part | No. of points | Best-fit azimuth | Deviation from cardinality (DFC) | Notes |
|---|---|---|---|---|---|---|---|
| Sphinx temple | North | Enclosure rock wall | | 6 | 89° 31´ | –0° 29´ | |
| **Sphinx temple** | **North** | **Main wall** | | **8** | **90° 22´** | **+0° 22´** | |
| **Sphinx temple** | **East** | **Main wall** | | **9** | **359° 58.8´** | **–0° 01.2´** | |
| Sphinx temple | South | Passage wall | Western | 5 | 94° 33´ | +4° 33´ | |
| Sphinx temple | South | SE corner chamber, south wall | | 4 | 92° 44´ | +2° 44´ | |
| Sphinx temple | South | SE corner chamber, north wall | | 4 | 92° 42´ | +2° 42´ | |
| Sphinx temple | West | SW corner chamber, west wall, outer face | | 4 | 1° 41´ | +1° 41´ | |
| Sphinx temple | West | SW corner chamber, west wall, inner face | | 3 | 2° 13´ | +2° 13´ | |
| Sphinx temple | West | Inner wall, outer face | Southern | 3 | 0° 04´ | +0° 04´ | |
| Valley temple | East | Foundation wall, base | | 8 | 89° 48´ | –0° 12´ | |
| **Valley temple** | **East** | **Main wall** | | **12** | **89° 37.0´** | **–0° 23.0´** | |
| Valley temple | South | Outer wall, outer face | | 10 | 90° 06.2´ | +0° 06.2´ | |
| **Valley temple** | **South** | **Main wall** | | **7** | **90° 32´** | **+0° 32´** | |
| Valley temple | West | Foundation wall | | 8 | 0° 44´ | +0° 44´ | |
| Valley temple | West | Outer wall, outer face | | 3 | 359° 31´ | –0° 29´ | |
| **Valley temple** | **West** | **Main wall** | | **6** | **0° 33´** | **+0° 33´** | |

To judge by the only 3-point alignment that we could identify, on the north side, the satellite pyramid of Khafre is oriented slightly clockwise at +0° 04´. Two additional pairs of points on short straight segments of the west wall yielded a deviation from cardinality of –0° 12´ (towards the southern end) and +0° 02´ (northern end) respectively, but such readings yield no error checks and can only be taken as generally affirming the +0° 04´ figure.

We measured a line of 8 points on the northern edge of Khafre's causeway, which yield our best estimate of its orientation (azimuth 103° 26´/283° 26´). The discrepancy between this and the figure obtained from 4 points on the southern edge (103° 33´/283° 33´) appears to be due to the poorer state of the latter. A 7´ discrepancy over the entire length of the causeway (495m) would mean a difference between the width at both ends of 1.01m, but existing plans (e.g. Maragioglio and Rinaldi 1966, pl. 5, fig. 1 & pl. 14, fig. 1) provide no indication that the two sides of the Khafre causeway are anything but parallel. Our measurements yield a mean width of 8.29m over the section surveyed.

The south and west walls of the valley temple display a consistent DFC close to +0° 32′, but that of the east wall is almost a degree different, at –0° 23′. This implies that the building is somewhat narrower at its northern end, but also that the mean north-south axis is very close to the true meridian, with a DFC of just +0° 05′.

The most reliable sets of data relating to the Sphinx temple were obtained from its east and north walls, yielding DFCs of –0° 01′ and +0° 22′ respectively. As it was not possible to obtain readings from the west wall, the east wall orientation—which is extremely close to true north-south—gives our best estimate of the axial orientation. A small set of readings from the southernmost segment of the western inner wall yielded a DFC value of +0° 04′, suggesting that the two long interior walls were also carefully aligned cardinally. A limited sample of readings from other internal structures gave much more disparate DFC values, between +1.5° and +3°.

It is evident that the passageway separating the two temples is skewed away from true east-west by several degrees, in the same sense as the Khafre causeway. Measurements of the southern wall of the Sphinx temple yielded a DFC of +4° 33′.

The data are summarised in Table 4b.

**Table 4b: Khafre's pyramid and associated structures: summary data**

| Structure | Approx. foundation date * | Deviation from cardinality (DFC) | | | | | |
|---|---|---|---|---|---|---|---|
| | | North side | East side | South side | West side | Mean (all sides) | Mean (E & W sides) |
| Main pyramid | 2556 BC | **–0° 04′** | **–0° 04′** | **–0° 06′** | **–0° 04′** | **–0° 05′** | **–0° 04′** |
| Satellite pyramid | 2556 BC | +0° 04′ | | | | | |
| Valley temple | 2556 BC | | –0° 23′ | +0° 32′ | +0° 33′ | | +0° 05′ |
| Sphinx temple | 2556 BC | +0° 22′ | –0° 01′ | (+4° 33′) | | | |
| Causeway | 2556 BC | +13° 26´ | N/A | +13° 33´ | N/A | +13° 29´ | N/A |

**Notes**

\* See note 7. The dates quoted here (purely for reference and not critical to the arguments presented) are arrived at by taking Shaw's (2000: 480) most likely date for Khafre's accession, 2558 BC, and then tentatively assuming that all the structures were started a couple of years into the reign.

### 3.5 *The pyramid of Menkaure*

Surveying this pyramid proved a challenge because its foundations were covered with rubble; and we were only able to survey the alignment of courses of stones on the pyramid itself.

Largely intact courses of casing stones are only found on the north and east sides. A few casing stones remained on the south side so we surveyed a combination of casing stones and coarse core blocks on the second and seventh courses respectively. On the west side no casing stones remain so we surveyed the extant raw core blocks on the ninth and eleventh courses.

**Table 5a: The pyramid of Menkaure: basic data**

| Side | Indication | Part | No. of points | Best-fit azimuth | Deviation from cardinality (DFC) | Notes |
|---|---|---|---|---|---|---|
| North | 4th course, outer edge | Western | 5 | 90° 22´ | +0° 22´ | |
| North | 6th course, outer edge | Central | 5 | 90° 20´ | +0° 20´ | |
| North | 7th course, outer edge | Central | 6 | 90° 18´ | +0° 18´ | |
| **North** | **7th course, etched line** | | 20 | 90° 18.5´ | +0° 18.5´ | |
| **North** | **9th course, outer edge** | | 22 | 90° 21.5´ | +0° 21.5´ | |

**Table 5a (ctd)**

| Side | Indication | Part | No. of points | Best-fit azimuth | Deviation from cardinality (DFC) | Notes |
|---|---|---|---|---|---|---|
| **East** | 3rd course, outer edge | Northern | 11 | 0° 06.5´ | +0° 06.5´ | |
| **East** | 6th course, outer edge | | 21 | 0° 16.5´ | +0° 16.5´ | |
| South | 2nd course, outer edge | Western | 3 | 89° 48.5´ | –0° 11.5´ | |
| **South** | 7th course, outer edge | | 26 | 90° 31.8´ | +0° 31.8´ | c |
| **West** | 9th course, outer edge | | 23 | 0° 29.5´ | +0° 29.5´ | c |
| **West** | 11th course, outer edge | | 21 | 0° 19.7´ | +0° 19.7´ | c |

**Notes**

c  Measurements taken from coarse blocks only, in the absence of casing stones

An issue to address at the outset is the concavity of the faces. Maragioglio and Rinaldi (1967: 36–37) have stated that, at least in the upper part of the pyramid, the extant casing is perfectly flat while the internal packing blocks formed a concavity in the center of each face:

> "The lower part of the filling of the steps, up to the height of the granite casing, in general follows an inclined surface like that of the outer faces of the pyramid. The thickness of the casing in this area, except for the bonding blocks, was thus fairly constant. The upper part, on the other hand, corresponding to the limestone casing, has on all of the faces—and particularly to the south—a marked concavity which seems intentional. The concavity appears greater in the lower part of this section and diminishes as it goes upwards. The thickness of the white limestone casing was thus greater towards the centre of the faces than towards the corners, as in Cheops" (Maragioglio and Rinaldi 1967: 36).

To see whether our data are consistent with Maragioglio and Rinaldi's statement we have taken each of the instances where we have readings from the entire length of a course and calculated the best-fit azimuth for each half as well as the whole. The data are presented in Table 5b.

**Table 5b: The Pyramid of Menkaure: concavity**

| | | | Mean deviation from cardinality (DFC) | | | |
|---|---|---|---|---|---|---|
| Wall | Course | Type | Whole length | 'CCW' half | 'CW' half | 'CCW' half – 'CW' half |
| North | 7th | Casing | **+0° 18´** | +0° 42´ | –0° 08´ | +0° 50´ |
| North | 9th | Casing | **+0° 22´** | +0° 57´ | –0° 04´ | +1° 01´ |
| East | 6th | Casing | **+0° 17´** | +0° 37´ | +0° 16´ | +0° 21´ |
| South | 7th | Internal | **+0° 32´** | +0° 34´ | –0° 02´ | +0° 36´ |
| West | 9th | Internal | **+0° 30´** | +1° 00´ | +0° 23´ | +0° 37´ |
| West | 11th | Internal | **+0° 20´** | +0° 13´ | +0° 13´ | 0° 00´ |

**Notes**

- 'CCW half' means the half of the wall in the counter-clockwise sense, i.e. the western half of the north wall, northern half of the east wall, etc., while 'CW half' is the half in the clockwise direction, i.e. the eastern half of the north wall, southern half of the east wall, etc.

The figures in the final column give the difference between the best-fit azimuths for the two halves of the segment in question. A positive value indicates that the segment in question is concave, as is the case with all the segments except the last.

Our data relate exclusively to the lower part of the pyramid, cased in granite (which extended up to the 16th course), and seem to demonstrate that it was not only the upper part of the pyramid that was concave. All sets of readings taken on the 6th to 9th courses show the sides in question to be concave. The west wall of the 11th course, on the other hand, runs in the same direction at both ends but changes direction in the central part. This is evident if we further divide the segment into quarters, whereupon the best-fit DFCs of the successive quarters running from south to north are +0° 01´, +0° 19´, +0° 29´ and +0° 08´. (This also explains how the DFC of the whole length is greater than the DFC of either half.)

We cannot say that the concavity diminishes in any even or consistent way from the 6th course up to the 9th. It would not be surprising that concavity was less important for stone courses further up the pyramid with less weight above them to support, but the only directly comparable readings on the same side (north wall, 7th and 9th courses) actually show that the higher course has greater concavity. Our results also suggest that there was increasing irregularity at these higher levels, at least among the internal packing stones.

Where we have data from casing and internal stones on the same courses (7th and 9th), it appears that the casing rather than the internal stones are more concave, implying that the concavity is emphasized rather than evened out by the casing, in contrast to Khufu's pyramid (and, according to M&R, to the upper levels of Menkaure). This would be surprising if, the reason for the concavity was merely to aid structural integrity and raises the possibility that there was also an aesthetic motive, as has been suggested by Maragioglio and Rinaldi (1965: 104) at Khufu.[8] However, we cannot be certain of this since there may be a difference from wall to wall, even in the same course.

In summary, while acknowledging the 'sparse' nature of our data we tentatively conclude that the 6th to 9th stone courses were concave, the difference in the orientation of two best-fit straight lines fitted to each half of the wall being of the order of a half to one degree. This would mean that the central part is approximately 0.4–0.8m closer to the centre of the pyramid than would be the case if the wall were flat.[9]

Regarding the orientation, our data also suggest that the distinction between facing and coarse blocks is not a critical one in estimating the intended orientation of the pyramid. Thus, excluding the anomalous value from the second course on the south side (given that it is based on only 3 points), we summarise all the orientation data in Table 5c. Evidently it is meaningless, in view of the quality of the data, to quote any values to better than the nearest arc minute.

---

[8] In their account of the Khufu pyramid, Maragioglio and Rinaldi (1965: 104) state : "Another possible explanation for the concave disposition of the nuclear courses could be [that] the builders wanted to give a curved form, if not to the casing, at least to the nucleus in order to prevent the faces from sliding, especially in the middle. … The hypothesis that the casing faces might have been slightly concave, although in less measure than the nucleus, should not be rejected either. Besides obtaining the practical result mentioned above, there would also have been an aesthetic one: the optic aberration would have been corrected, the pyramid edges made sharper and more defined and the faces would have appeared to be flat, without considering that the eventual errors in dressing the faces would have been disguised better."

[9] We emphasize that this conclusion is based on two best-fit straight lines fitted to each half of the wall, and would need to be modified if the actual wall were in fact curved.

**Table 5c: The Pyramid of Menkaure: summary data**

| Wall | Mean deviation from cardinality (DFC) | | | | | | |
|---|---|---|---|---|---|---|---|
| Course | 3rd | 4th | 6th | 7th | 9th | 11th | Mean |
| North |  | +0° 22´ | +0° 20´ | +0° 18´ | +0° 22´ |  | **+0° 20´** |
| East | +0° 07´ |  | +0° 17´ |  |  |  | **+0° 12´** |
| South |  |  |  | +0° 32´ |  |  | **+0° 32´** |
| West |  |  |  |  | +0° 30´ | +0° 20´ | **+0° 25´** |
|  | +0° 07´ | +0° 22´ | +0° 18´ | +0° 24´ | +0° 26´ | +0° 20´ |  |

Despite the sparse nature of the data that we were able to retrieve, we can certainly confirm that the Menkaure pyramid is skewed clockwise from cardinality, unlike those of Khufu and Khafre. As far as can be judged from the courses of stones, the mean DFC of the sides is +0° 22´; which is also the mean DFC of courses 4 to 11 (excluding course 3). So this forms our best estimate of the intended orientation.

## 3.6 Structures associated with Menkaure's pyramid

The basic data from the pyramid complex of Menkaure are presented in Table 6a. All measurements at MQ1–MQ3 refer to casing stones.

**Table 6a: Pyramid of Menkaure, associated structures: basic data**

| Structure | Side | Indication | No. of points | Best-fit azimuth | Deviation from cardinality (DFC) | Notes |
|---|---|---|---|---|---|---|
| Queen's pyramid MQ1 | North | 6th course, outer edge | 6 | 88° 59´ | –1° 01´ |  |
| Queen's pyramid MQ1 | East | 6th course, outer edge | 4 | 1° 42´ | +1° 42´ |  |
| Queen's pyramid MQ1 | East | 6th course, outer edge | 3 | 90° 30´ | +0° 30´ |  |
| Queen's pyramid MQ1 | South | 6th course, outer edge | 3 | 90° 54´ | +0° 54´ |  |
| Queen's pyramid MQ1 | West | 6th course, outer edge | 4 | 0° 31´ | +0° 31´ |  |
| Queen's pyramid MQ2 | North | 3rd course, outer edge | 5 | 90° 27´ | +0° 27´ |  |
| Queen's pyramid MQ2 | West | 7th course, outer edge | 4 | 0° 12´ | +0° 12´ |  |
| Queen's pyramid MQ3 | North | 1st course, bottom | 6 | 90° 39´ | +0° 39´ |  |
| Queen's pyramid MQ3 | East | 1st course, bottom | 3 | 359° 17´ | –0° 43´ |  |
| Queen's pyramid MQ3 | West | 1st course, bottom, Southern part | 3 | 2° 39´ | +2° 39´ |  |
| Causeway | North | Outer edge of pavement | 6 | 89° 10´ | –0° 50´ |  |

The main issue in relation to the queens' pyramids is the extent to which the scattered DFC values simply reflect their poor state of preservation or may be due in part to a lack of precision in their original construction. Certainly, the data from MQ1 tell us very little except that this pyramid was approximately oriented cardinally to within a couple of degrees.

On the other hand, there is a degree of consistency between the DFC values obtained from MQ2 and MQ3 (excluding lines of only 3 points) which seems to imply that these pyramids generally followed a well defined practice of orientation resulting in DFC values (between +0° 12´ and +0° 39´) not dissimilar to Menkaure's pyramid itself. The DFC value obtained from the west side of MQ1, at least, is consistent with this.

A small segment of well-preserved pavement edge on the northern side of the Menkaure causeway yielded an azimuth of 89° 10´ (DFC –0° 50´). Thus Menkaure's causeway, although approximately cardinal (unlike the causeways of Khufu and Khafre), deviates from cardinality by a significant amount and in the opposite sense to the pyramids of Menkaure and his queens.

These data (excluding lines of only 3 points) are summarised in Table 6b.

**Table 6b: Menkaure's Pyramid and associated structures: summary data**

| Structure | Approx. foundation date * | Deviation from cardinality (DFC) | | | | | |
|---|---|---|---|---|---|---|---|
| | | North side | East side | South side | West side | Mean (all sides) | Mean (E & W sides) |
| Main pyramid | 2530 BC | **+0° 20′** | **+0° 12′** | **+0° 32′** | **+0° 25′** | **+0° 22′** | **+0° 18′** |
| Satellite pyramid MQ1 | 2530 BC | –1° 01′ | +1° 42′ | | +0° 31′ | | +1° 11′ |
| Satellite pyramid MQ2 | 2530 BC | +0° 27′ | | | +0° 12′ | | |
| Satellite pyramid MQ3 | 2530 BC | +0° 39′ | | | | | |
| Causeway | 2530 BC | –0° 50′ | N/A | | N/A | | N/A |

**Notes**
\* See note 7. The dates quoted here (purely for reference and not critical to the arguments presented) are arrived at by taking Shaw's (2000: 480) most likely date for Menkaure's accession, 2532 BC, and then tentatively assuming that all the structures were started a couple of years into the reign.

In addition, alignment data for Menkaure's valley temple—usually buried by sand but cleared during excavation—have been provided courtesy of the GPMP. They give the north-south alignment as 359.8°/179.8° and the east-west alignment as 90.2°/270.2°.

# 4 Comparative analysis

In this section we compare our 2006 survey results with those published by other authors. The surveys conducted by Cole, Dorner, and Lehner and Goodman were restricted to the pyramids of Khufu and Khafre, so full comparative analyses are only possible in the case of these two pyramids.

## *4.1 Pre-Giza pyramids: the Third and Early Fourth Dynasties*

As a preliminary, we present and review alignment data collected by other authors from pre-Giza pyramids in dynasties 3 and 4. Included in this review are the 'Bent Pyramid' and 'Red Pyramid' (the south and north pyramids, respectively) at Dashur, both built by Snefru early in dynasty 4 and located approximately 40 km south of Giza (Reisner 1942: 1-3). We also include the Meidum pyramid, about 100km south of Giza, whose construction was also completed in Snefru's reign although it has been speculated that it was begun by Huni (Snefru's father) (Stadelmann 1980; Malek 2000: 93). Of the other ten pyramids two, belonging to Djoser and Sekhemkhet respectively, are located at Saqqara (between Dashur and Giza); and one at Zawiyet el-Aryan was (possibly) built by the pharaoh Khaba (Lehner 1997: 95). The other seven are small step pyramids located between the Fayoum and Elephantine Island (the southern border of upper Egypt) (*ibid*.). The function of these provincial step pyramids is unknown because they do not have burial chambers or (except in Seila's case) associated chapels (Lehner 1997: 96).

The published data are summarized in Table 7. We have retained the level of precision quoted in the original publications, although in the case of the three values quoted to the nearest arc second it is clear from the level of discrepancy between these and independent measurements of the same structures that they are unlikely to be accurate to better than a few arc minutes. In the case of El Kula and Seila, there is a gross discrepancy between the two independent readings.

**Table 7: Pre-Giza pyramid alignments**

| Location | Pharaoh | Approx dates of reign* | Deviation from cardinality | Source | Notes |
|---|---|---|---|---|---|
| Saqqara | Djoser | 2667–2648 | +3° | Lauer 1960: 99; Romer 2007: 279 | |
| Saqqara | Sekhemkhet | 2648–2640 | –11° | Lauer 1960: 99; Romer 2007: 279 | |
| Zawiyet el-Aryan (Layer pyramid) | ?? | ?2640–2613 | –9 to –8° | Lehner 1996: 510; Romer 2007: 279 | a |
| Elephantine | ?Huni | 2637–2613 | –17° | Lehner 1997: 96; Belmonte & Shaltout 2009: 312, appendix II | |
| Edfu (El Ghoneimiya) | ?? | ?2640–2613 | +3.5° | Belmonte & Shaltout 2009: 312, appendix II | a, b |
| El Kula | ?? | ?2640–2613 | –43.75° | Belmonte & Shaltout 2009: 312, appendix II | a, c |
| Ombos/Naqada | ?? | ?2640–2613 | +24.5° | Belmonte & Shaltout 2009: 312, appendix II | a, d |
| Sinki | ?? | ?2640–2613 | –41.25° | Belmonte & Shaltout 2009: 312, appendix II | a, e |
| Zawiyet el-Meitin (Layer pyramid) | ?? | ?2640–2613 | –19° | Belmonte & Shaltout 2009: 312, appendix II | a, f |
| Seila | Snefru | 2613–2589 | 0° | Belmonte & Shaltout 2006: 179–181 | g |
| Meidum | Snefru | 2613–2589 | –0° 18′±2′ | Belmonte 2001: S2 | |
| Dashur, south ('Bent Pyramid') | Snefru | 2613–2589 | –0° 12′±2′ | Belmonte 2001: S2 | |
| | | | –0° 12′±2′ | Belmonte 2001: S2 | |
| Dashur, north ('Red Pyramid') | Snefru | 2613–2589 | –0° 5′ 0″ | Isler 2001: 158. Measured by Joseph Dorner | |
| | | | –0° 9′±2′ | Belmonte 2001: S2 | |

**Notes**

\* See note 7. The dates quoted here (purely for reference and not critical to the arguments presented) are those of the pharaoh's reign taken from Shaw (2000: 480).

a It is uncertain for which pharaoh this pyramid was built. The date range given covers the reigns of Khaba (2640–2637 BC) and Huni (2637–2613 BC), again according to Shaw (2000: 480), but this assignation is far from certain.

b Lehner (1997: 96) lists the DFC as ~ 0°, which is broadly consistent with Belmonte and Shaltout's more accurate value.

c   Lehner (1997: 96) lists the DFC as 0° but this seems to refer to the diagonal, which is close to cardinal.
d   Lehner (1997: 96) lists the DFC as ~ –12°, which is inconsistent with Belmonte and Shaltout's value.
e   Lehner (1997: 96) lists the DFC as ~ 0° but this seems to refer to the diagonal, which is close to cardinal.
f   Lehner (1997: 96) lists the DFC as ~ –20°, which is consistent with Belmonte and Shaltout's more accurate value.
g   Lehner (1997: 96) lists the DFC as –12°, which is inconsistent with Belmonte and Shaltout's value.

Based on these data, it is clear that the practice of orientating royal pyramids cardinally with remarkable precision was well established during Snefru's reign. It is also possible that the orientations of Meidum, The Bent Pyramid at Dashur, and the Red Pyramid at Dashur represent successive refinements getting ever closer to true cardinality, while always deviating counterclockwise, as argued by Spence (2000). The data from Seila are less accurate, and this pyramid is perfectly cardinal within the larger measurement errors.

In contrast, the data from the earlier pyramids, and particularly from the small step pyramids, really show no obvious trends at all.

## *4.2   The Great Pyramid of Khufu*

Four noteworthy surveys of the Great Pyramid have been conducted prior to the one reported in this paper: those of Petrie (1883), Cole (1925: 1–9), Dorner (1981) and Lehner and Goodman (1984; Lehner 1985a). As already noted, Dorner's survey combined his own theodolite measurements with distance measurements taken from Cole.

The results of all five surveys are summarized in Table 8. For clarity, earlier survey data, originally expressed in degrees, minutes and seconds, have been converted to degrees, minutes and decimals of minutes.

**Table 8: The Great Pyramid of Khufu: comparative data**

| Side | Petrie (1883: 39) | Cole (1925: 6)[10] | Dorner (1981: 77) | Lehner and Goodman (1984) | Nell & Ruggles | **Mean deviation from cardinality (DFC)** |
|---|---|---|---|---|---|---|
| North | –0° 03.3′ | –0° 02.5′ | –0° 02.5′ | –0° 03.1′ | –0° 03.6′ | **–0° 03.0′** |
| East | –0° 04.0′ | –0° 05.5′ | –0° 03.4′ | –0° 03.7′ | –0° 03.4′ | **–0° 04.0′** |
| South | –0° 03.7′ | –0° 02.0′ | –0° 02.5′ | +0° 00.5′ | –0° 00.5′ | **–0° 01.7′** |
| West | –0° 03.9′ | –0° 02.5′ | –0° 02.8′ | –0° 07.1′ | –0° 03.7′ | **–0° 04.0′** |
| **Mean deviation from cardinality** | **–0° 03.7′** | **–0° 03.1′** | **–0° 02.8′** | **–0° 03.4′** | **–0° 02.8′** | **–0° 03.2′** |

These values are broadly consistent, but with some obvious discrepancies such as Lehner and Goodman's value for the west side and Cole's for the east side, and the fact that Lehner and Goodman's and our values for the south side are somewhat higher than those of the three earlier surveyors. The obvious possibility is that some if not all of these discrepancies are due to different authors measuring different things.

---

[10]   "East" and "south" have been mistakenly transposed in the comparative table at the foot of page 7.

Lehner and Goodman (1984) measured corner-to-corner alignments between the brass markers (see Lehner 1985a: 27) and their azimuth values for the north and west sides concur to within 00.5′ and 00.2′ respectively with ours (Table 1a). Lehner and Goodman's value of –7.1′ for the west side and ours of –7.3′ for the azimuths between the corner markers are also in broad agreement with Petrie's estimate of –7.5′ for the rough core masonry, listed in a separate table (Petrie 1883: 38).

Petrie (1883: 37) is explicit that he measured the orientation of the casing [foundation] base, which he found to be significantly more regular than the core masonry (ibid.: 40). Cole (1925: 5) measured the top edge of the ground casing blocks on the south side and the bottom of casing blocks at the pavement on the other sides. Thus their values (and Dorner's) should be directly comparable with ours as listed in Table 8, since all are based upon measurements of the casing foundation. Leaving aside the south side, our values concur with Petrie's to within 0.6′ and to the other authors' to within 2.1′. Our value for the south side is significantly different from that of the three earlier authors, differing from Petrie's value by 3.2′. It may be that improvements in technology, and in particular the ability to take multiple readings along alignment of points conveniently using EDM devices and undertaking best-fit analyses, can explain some of these differences.

The data certainly confirm that the largest pyramid in the history of pyramid building was also the most accurately cardinally aligned (Lehner 1997: 108), none of the four sides deviating by more than ca. 4 arc minutes in the anticlockwise sense. Khufu's pyramid, built approximately 40 km north of Dashur, appears to mark the culmination of a process of refinement in determining the orientation that began in the era of Snefru, his father.[11] This analysis does, however, cast doubt upon whether the precision to which the relevant azimuths are quoted in the literature (often to the nearest arc second) matches their accuracy.

The data appear to sustain the following conclusions:
- It is probably misleading to quote the orientations to a precision of better than 1′.
- The mean deviation from cardinality is between –4′ and –3′, consistent with Petrie's best estimate of –3.7′.
- The easterly and westerly orientations (north and south walls) are slightly closer to true cardinality than those in the north-south directions.

It is tempting, on the basis of the mean deviations from cardinality in Table 1b, to conclude that the east and west sides are closer to parallel with each other than are the north and south sides. Indeed, our own data suggest that the south side is out of kilter with the other three, actually being oriented to within 1′ of true east-west. However, in view of differences between the different sets of survey results it is not possible to make these assertions with absolute confidence.

The data also highlight potential differences between the foundations and the stone courses, echoing Petrie's opinion (1883: 126) that "the difference of the casing and core azimuths of the Great Pyramid [averaging –3.7′ and –5.3′ respectively] shows that probably a re-determination of the N. was made after the core was finished; and it must be remembered that the orientation would be far more difficult to fix after, than during, the construction; as a high face of masonry, for a plumb-line, would not be available."

## 4.3   Khufu's queens' pyramids

According to our survey KQ2 is oriented with a DFC of –0° 15′. This accords with Petrie's (1883: 121) estimate for the passage of KQ1 (DFC –0° 15′±2′), but not with his value for the azimuth of the passage in KQ2, which he determined to be –0° 3′±1 (*ibid.*: 123).

---

[11] Petrie noted, furthermore, that the inner corridors of Khufu's pyramid were also closely aligned to the cardinal directions, the mean axis of the main passage being –3.7′ (whole length) or –5.8′ (built part alone) (Petrie 1883: 58).

## 4.4 The Pyramid of Khafre

Petrie (1883), Dorner (1981) and Lehner and Goodman (1984; Lehner 1985a; 1986) have undertaken independent surveys of Khafre's pyramid. The results of these plus our own surveys are summarized in Table 9. For clarity, earlier survey data, originally expressed in degrees, minutes and seconds, have been converted to degrees, minutes and decimals of minutes.

**Table 9: Khafre's pyramid: comparative data**

| Side | Petrie (1883: 97) | Dorner (1981:80–81) | Lehner (1986: 31) | Nell & Ruggles | **Mean deviation from cardinality (DFC)** |
|---|---|---|---|---|---|
| North | –0° 05.5′ | –0° 05.2′ | –0° 04.6′ | –0° 03.8′ | **–0° 04.8′** |
| East | –0° 06.2′ | –0° 06.0′ | –0° 05.3′ | –0° 04.0′ | **–0° 05.4′** |
| South | –0° 05.7′ | –0° 05.7′ | –0° 03.5′ | –0° 05.8′ | **–0° 05.2′** |
| West | –0° 04.4′ | –0° 06.0′ | –0° 06.3′ | –0° 04.2′ | **–0° 05.2′** |
| **Mean deviation from cardinality** | **–0° 05.4′** | **–0° 05.7′** | **–0° 04.9** | **–0° 04.5′** | **–0° 05.1′** |

As in the case of Khufu's pyramid these data are broadly consistent, although our survey yields the estimate of the mean DFC closest to true cardinality. Our data are in close agreement with Petrie's—who also measured the casing foundation base (1883: 30–31)—on the south and west sides but yield a higher (more clockwise) DFC than all three other surveys on the north and east sides. In broad terms the accumulated data support the long-known conclusion that Khafre's pyramid is not quite as well cardinally oriented as Khufu's, deviating further from cardinality than the Great Pyramid by somewhere between 1.5′ and 2.0′.

Taken at face value, the data reinforce the conclusion reached at Khufu's pyramid that it is probably misleading to quote the orientations to a precision of better than 1´. They also appear to sustain the following conclusions:
- The mean deviation from cardinality is not significantly greater than 5´, and possibly somewhat less, so that the pyramid is slightly better aligned to the cardinal directions than Petrie and Dorner estimated.
- While our own data suggest, as at Khufu's pyramid, that the south wall is slightly out of kilter with the other three, but the other surveys do not substantiate this.

Lehner and Goodman's data relate to the postholes surrounding Khafre's pyramid, which are better defined than those around Khufu's (Lehner 1985a: 31; see Fig. 6). Upon comparing their posthole survey data with the Khafre pyramid data collected by Dorner they conclude that the lines of postholes on all but the west side are better oriented to the cardinal directions than the sides of the pyramid itself (Lehner 1985a: 37; 1986: 32), which supports the opinion that in setting out the pyramids the ancient Egyptians used survey posts and cords, and the postholes are remnants of this process. Our data suggest that the casing foundation base was better cardinally oriented than the posthole lines on all but the south side.

## 4.5 The pyramid of Menkaure

Petrie surveyed Menkaure's pyramid but, on concluding that the casing pavement was never finished, took measurements of the end points of the best preserved granite courses on each side. He then calculated the corner-to-corner azimuth of each side by 'projecting down' the corners to a common level assuming a slope of 51° (omitting the west side, where he could not obtain the requisite data from the north end) (Petrie 1883: 111–112). We summarize Petrie's data in Table 10. He also gives the azimuth of the passage as +0° 13.3´ (ibid.: 117).

**Table 10: Menkaure's pyramid: comparative data**

| Side | Petrie (1883: 111) | Course |
|---|---|---|
| North | +0° 16.8′ | 4th (W end); 1st (E end) |
| East | +0° 12.4′ | Casing foundation base (N end); "foot of rough casing" (S end) |
| South | +0° 13.0′ | 3rd (E end); 1st (W end) |
| West | — | |
| **Mean deviation from cardinality** | **+0° 14.1′** | |

Dorner did not carry out his own survey of Menkaure's pyramid, merely quoting Petrie's results (Dorner 1981: 82).

While we did obtain a DFC of +0° 07′ from the 3rd course on the east side, our DFC values obtained from higher stone courses are between +0° 17′ and +0° 32′.

Petrie's data are not directly comparable with ours (Tables 5a–5c), but the two sets of data perhaps complete a picture in which the higher levels were 'twisted round' with respect to the foundation, a possibility that Petrie himself seems to have ruled out.[12]

## 5   Discussion

### *5.1   Patterns of orientation*

In order to examine some of the broader questions relating to the motives for orienting the various structures on the Giza plateau and the techniques used, it is useful to summarise the overall orientation trends that emerge from all the data discussed so far. Table 11 represents an attempt to summarize these trends. Note that columns 3 and 4 of this table refer to the N-S and E-W *directions*, the N-S direction being that indicated by the alignments of the east and west sides or walls of the structures concerned, and the E-W direction being that indicated by the north and south sides or walls.

**Table 11: Overall patterns of orientation: summary data**

| Approx date* | Structure | Mean deviation from cardinality (DFC) | | Notes |
|---|---|---|---|---|
| | | **N-S direction** | **E-W direction** | |
| Khufu complex | | | | |
| 2587 BC | **Main pyramid** | **–0° 04.0′** | **–0° 02.4′** | A |
| 2587 BC | Mortuary temple | 0.0° | | |
| 2585 BC | Queen's pyramid KQ2 | –0.3° | | |
| 2585 BC | Western cemetery | –2.3° | (–2.6°) | b |
| 2535 BC | Southern cemetery | **–0.6°** | **(–0.6°)** | cd |
| 2583 BC | Eastern cemetery | –0.5° to +1.0° | | |

---

[12] "It must be remembered that if any different base level should be supposed to have been intended, it will make no difference in the above azimuths, nor in the differences between the sides." (Petrie 1883: 212).

**Table 11 (ctd)**

| Approx date* | Structure | Mean deviation from cardinality (DFC) | | Notes |
| --- | --- | --- | --- | --- |
| | | N-S direction | E-W direction | |
| Khafre complex | | | | |
| 2556 BC | **Main pyramid** | **–0° 05.3′** | **–0° 05.0′** | E |
| 2556 BC | Satellite pyramid | | *+0.1°* | f |
| 2556 BC | Causeway | N/A | **+13.3°** | |
| 2556 BC | Valley temple | **+0.1°** | +0.5° | |
| 2556 BC | Sphinx temple | ..0.0° | +0.4° | g |
| Menkaure complex | | | | |
| 2530 BC | **Main pyramid** | **+0° 18′** | **+0° 26′** | |
| 2530 BC | Queen's pyramid MQ1 | +1.2° | –1.0° | |
| 2530 BC | Queen's pyramid MQ2 | +0.2° | +0.5° | |
| 2530 BC | Queen's pyramid MQ3 | | +0.6° | |
| 2530 BC | Causeway | N/A | –0.8° | |
| 2530 BC | Valley temple | –0.2° | +0.2° | h |

**Notes**
* See note 7 and Tables 2b, 4b, 6b.
- Bold face indicates a figure derived as the mean of measurements on opposite sides; otherwise the figure is from one face only. The alignments of shorter sides of mastabas in the Khufu complex, less likely to be significant (or precisely defined or accurately measured) in themselves, are shown in brackets. Readings considered less reliable (reasons given in the notes) are italicized.
A  The figures obtained from our data alone are –0° 03.6′ and –0° 02.1′ respectively.
b  Based solely on averages of measurements of straight segments with only three surveyed points (see Table 1a).
c  Measurements are from the mastabas in the western half of the cemetery only.
d  The five measurements from shorter sides (Table 2b) yield a mean DFC in the east-west direction of –0.6°, but the wide scatter of these readings suggests that they merely reflect a broad practice of constructing the shorter walls to be broadly perpendicular to the longer ones, with no great precision.
E  The figures obtained from our data alone are –0° 04.1′ and –0° 04.8′ respectively.
f  Based on the measurement of a single straight segment with only three surveyed points (see Table 4a).
g  Data from the skewed south (passage) wall (see Table 4b) have been excluded.
h  GPMP data

Our own data from the three main pyramids, and the comparative analysis with earlier surveys, reveals a level of complexity in determining the intended orientations that is not always acknowledged in discussions of the factors that may have influenced those intended orientations. Certainly, azimuths quoted to second-of-arc precision are completely unjustified, and decimals of a minute of arc may be misleading. Even minute-of-arc precision may be unjustified in the case of Menkaure's pyramid, given that relevant data can only be obtained at the present time from stone courses, not casing foundations.

That said, our data support the widely quoted assertion that the mean orientation of Khafre's pyramid deviates slightly more from true cardinality than does Khufu's. Nonetheless, according to our own data, the difference between the north-south orientation of the two pyramids is only 0.5′. Khufu's pyramid is some 1.5′ to 2′ off perpendicular, much more so than Khafre's, and the north and south sides are significantly non-parallel. Yet it is the east-west direction that is closer to true cardinality, and our data concur with Lehner and Goodman's—though not to the earlier surveys—in singling out the south wall as the one closest of all to true cardinality, deviating by no more than half a minute (see Table 8).

As we have already noted, the orientations of the Khufu and Khafre pyramids appear to follow a precedent for anti-clockwise orientation already evident during the earlier part of dynasty 4, and likely refined over time so as to be ever closer to true north. Menkaure broke with this tradition. Not only was his pyramid oriented clockwise, but it also deviated significantly more from true cardinality. While this could be because Menkaure's surveyors were less precise than those of Khufu and Khafre, the fact that his pyramid was constructed no less finely than the others (despite being significantly smaller with its facing never completed) suggests otherwise.

What about the 'bigger picture' provided by all the other structures on the plateau? Starting with Khufu's complex, there was a marked difference in orientation between his pyramids—his own and those of his queens—(DFC between –0.3° and 0.0°) and the burial structures of his sons and advisors in the surrounding cemeteries (DFC [WC] ≈ –2.3°; DFC [SC] ≈ –0.6°; DFC [EC] between –0.5° and +1.0°). While this could be due to poorer techniques being applied to ensure the cardinal orientation of the smaller tombs, two factors at least suggest otherwise. The first is a systematic increase in the DFC (orientation change in the clockwise sense) from west to east, the value being c. –3.0° in the case of the westernmost mastabas and c. +1.0° in the case of the easternmost ones. If a similar method was used to align all of these structures, this might give an insight into that method. The second factor is that the successive rows in the eastern cemetery—while generally increasing in DFC from west to east (the DFCs for the first four rows being –0.4°, –0.5°, +0.3° and +1.0°—see Table 2b)—switch from orientation more anticlockwise than Khufu's pyramids to orientation more clockwise. While it is tempting to suggest that the complete avoidance of the –0.3° to 0.0° range could mean that a deliberate attempt was made to avoid replicating the precise orientation of the larger pyramids, the evidence is too thin to give serious support to this speculation.

The available data from Khafre and Menkaure complexes offer a less complete picture. Khafre's Valley Temple and the Sphinx Temple are both aligned very closely to the north-south meridian, deviating, if anything, clockwise rather than anticlockwise. As far as can be ascertained, Menkaure's satellite pyramids seem to follow the clockwise orientation pattern of his own pyramid.

The alignment of the causeways of Khufu and Khafre, although not that of Menkaure, differ greatly from all other monuments on the Giza Plateau, and may offer some useful insights as a result. Khafre's causeway, as already discussed, is oriented with a DFC of between +13° and 14°, while that of Khufu is around –14° to –15° (Lauer 1947: 246; Goyon 1969; Lehner 1985b: 118–119). The approximate symmetry with respect to due east-west is noteworthy although not necessarily an end in itself for the builders. The skewing of Khafre's causeway is more likely to have arisen as a consequence of other considerations, for example ensuring that the axis of the Sphinx temple, the south side of the Sphinx, and the south side of Khafre's pyramid were in alignment (Lehner 1997: 129–130).

### 5.2   *The question of the primary direction: planning and construction at Giza*

Lehner (1997: 107) argues that construction on the Giza Plateau was carefully planned, so that it would fulfil the needs of more than one pharaoh, even before Khufu's pyramid began to be constructed. This idea is supported by the fact that the location of Khufu's pyramid was as far north as was possible on the Mokkatam limestone ridge, thereby leaving room for the construction of the other two pyramids. It is also supported by various evident alignments among the monuments themselves: for example, the eastern sides of Khafre's and Menkaure's Mortuary Temples are aligned respectively with the western sides of Khufu's and Khafre's pyramids. Jeffreys (1998) and Magli (2009a) extend this 'master plan' concept, suggesting that a diagonal line running through the south-eastern corners of all three main pyramids (the 'Giza diagonal') possibly deliberately pointed directly to the ancient city of Heliopolis (*Iunu*) some 20km to the north-east, but there are several problems with this hypothesis.[13]

After the site had been selected and the area leveled, a plan certainly had to be implemented that dealt not only with the orientation and internal organization of each structure on the plateau but also with the placement of the various structures in relation to one another.

According to Romer (2007: 331), both Petrie and the early-20th century Egyptologist Ludwig Borchardt argued that an east-west baseline for Khufu's pyramid was the first to be laid out, and the fact that the orientation of the casing foundation edge on the south side is closest of all to cardinality (Tables 1a and 1b) seems to support this idea. However, the archaeological evidence seems to argue to the contrary. Surrounding the Great Pyramid are a number of postholes, unfortunately in a poor state of preservation, that may well have been used by the ancient surveyors to insert measuring posts or gnomons (Clarke and Engelbach 1930: 125ff; Isler 2001: chs. 11 and 12; Romer 2007: 333–334; Lehner 1985c: 51–52). Most postholes can be seen nowadays on the east side of the pyramid, with many on the north side as well, but relatively few on the west and south sides. There are also many postholes close to the NW and NE corners but few around the SE and SW corners. Many more may originally have existed on the south side (where very little pavement remains) and been destroyed when the southern mastabas were erected during Khafre's reign, as well as during the recent construction of a boat museum.[14]

In addition to the rock-cut holes on the eastern side of the pyramid, there are several 'blank' corridors cut into the rock, which Petrie (1883: 50–51) refers to as "Trial Passages", noting that their design replicated many of the junctions and doorways of the interior corridors. The Trial Passages, and a trench that appears to be associated with them, run parallel to the pyramid's eastern baseline—i.e., they have a north-south orientation. A row of rock-cut holes that crosses both the trench and the Trial Passages appears to be aligned to structural elements inside the Great Pyramid (Lehner 1985c: 51–52; figs 9 and 12). Lehner dates the trench to the construction of the pyramids and judges that these postholes were probably cut at the same time. All this suggests that the eastern side of the Great Pyramid was the primary focus of pyramid construction and survey and that the primary baseline was north-south, with offsets being used to mark out the inner passages of the pyramid (cf. Romer 2007: 346–9), although the corners were evidently important as well.

If, as seems probable, many of these holes held stakes marking temporary points and directions during the process of determining the final alignment, then it seems likely that the primary astronomical sighting method used to set up the survey grid was upon circumpolar stars to the north. However, even if the north-south direction was the primary one, this process could well have involved more than one of the astronomical methods, one being used to check another. We cannot rule out the possibility of the use of noonday shadows, or the rising sun in the east. Unfortunately, the layout of the extant postholes is uninformative: many of those on the eastern side of the Great Pyramid appear to be randomly placed, and while the eye can pick out 'straight line' patterns among those on the northern side, running broadly parallel to the side of the pyramid, they do not in themselves define an obvious baseline direction.

The postholes surrounding Khafre's pyramid (Fig. 6) are more consistently placed. There are three parallel rows of postholes on each of the four sides, the two rows furthest from the pyramid being located on the outside limit of the enclosure wall and the third row along the inner line of the enclosure wall (Lehner 1985a: 32). On the north, east and south sides they run the entire length from corner to corner but on the west side they stop some 10m short of the corners. This suggests that the primary baseline might have been laid out using astronomical methods on any one of the north, east or south sides, the entire grid being established by perpendicular offset and/or the use of other astronomical methods. The fact that the holes on the east side are 9.1 to 9.9 m from the pyramid while those on the south, west, and north sides are 13.5 m from the pyramid could be taken to imply that the eastern baseline was special in some way and hence was the one laid out first. On the other hand, it is possible that the eastern postholes were simply placed closer to the pyramid to accommodate the mortuary temple, located close to Khafre's pyramid on its eastern side (ibid.: 33).

---

[13] Magli has suggested that the Giza plateau pyramids, as well as several pyramids at Saqqara, Abu Sir, Abu Ghurob, Zawiyet el-Aryan and Abu Roash were all symbolically connected to Heliopolis, home of the sun cult, through the alignments of their corners. However, there are various problems with this idea. For one thing, the view of Heliopolis from Abu Sir was obstructed: those standing at Abu Sir could not even see the 'Sun City' (Magli 2010a: 2235). For another, the putative alignment of pyramids at Saqqara involves different corners—the NW corners of Teti's and Semerkhet's pyramids, the SE corners of Userkaf's and Djoser's pyramids, and a complete diagonal of Unas' (Lehner 1985d; Magli 2010b, 3)—raising familiar problems of selection.

[14] They may also have been destroyed when rocks from quarries further to the south were delivered on the south side (Romer 2007: 329; Lehner 1985a: figs. 2 and 3C).

As its foundations have not been exposed for over a hundred years, it is unclear whether Menkaure's pyramid was also surrounded by survey postholes, but it is reasonable to assume that such postholes may exist.

In sum, while the balance of the evidence favours the hypothesis that the primary baseline faced north-south and was set out on the eastern side of each of the main pyramids-to-be, all possibilities remain open.

## *5.3    A context for orientation: solar cult*

Despite doubts about whether the orientation of the 'Giza diagonal' towards Heliopolis was actually deliberate, it is worth noting as a possible indication of a strong connection with sun worship, since Heliopolis, the most important city in ancient Egypt around 2600 BC, was also the home of the sun cult.

Although (Khafre's elder brother) Djedefre's pyramid at Abu Roash is small and poorly preserved, as discussed above, he is notable in being the first pharaoh to add the "Son of Re" (*sare*) name to the royal system of titles. The cult of the sun had certainly been important in earlier dynasties (David 1981; Quirke 2001; Krauss n.d.) but the addition of the name *sare* underscores a change in religion and the growing importance of the sun as a god to the ancient Egyptians. At this time, the god Horus (*hor*), whose name had long been a royal title, received a new association and gained great importance as the son of Re. In addition, *sare* and one other name of the royal titulature—"He of the Sedge and the Bee" (*nesw bit*)—were surrounded by a cartouche that was actually an elongated *shen* hieroglyph; a symbol comprising a circle of rope tied together at the bottom. The word *shen* derives from the verb *shenu*, which not only means to encircle and protect but may also have represented the sun surrounded by the universe (Wilkinson 1992: 193; Shaw and Nicholson 1995: 267). The first pharaoh who used the *shen* ring to encircle his names was Huni, the father of Snefru and grandfather of Khufu. It has been suggested (though there is no firm proof) that Huni started to build the second large step pyramid at Meidum (it was finished by Snefru, to whom its construction is generally attributed).

The use of the *shen* ring (evolving into the cartouche), and the incorporation of Re's name into the names of pharaohs from Khafre and Menkaure onwards, highlights the importance of the connection between the king and the universe and underscores Re's serious impact on Old Kingdom culture. The names "He of the Sedge and the Bee" and "Son of Re" became the most important names in the royal titulature and went on to be used throughout the history of Egypt (Wilkinson 1992: 195).

It has been argued for many years that the Sphinx and its Temple were constructed to promote the solar cult within Khafre's pyramid complex (Anthes 1971: 52–57). A more recent proposition—given that the Sphinx's face, placed on top of a lion's body, may be based on Khufu's image rather than Khafre's (Stadelman 1985: 124–126; 2003)—is that the cult in question was a Khufu cult: but this does not preclude the association with the sun. Lehner (1997: 129–131) has suggested that the Sphinx represented the pharaoh Khafre in his Horus form as the son of Re, that he crouches before the solar temple dedicated to Re, and makes obeisance to the sun god Re as he is born (sunrise) everyday on the eastern horizon. As he has observed (pers. comm., 2005), the Sphinx itself is surrounded by a giant cartouche. Khafre's Valley Temple seems to have had a connection with the day-night cycle, its upper level being associated with daytime and the sun, while the lower level is thought to be associated with the *Duat* (underworld): twenty-four statues placed there may have represented the 24 hours of the day and night (Ricke 1970; but see Lehner 1997: 126).

In view of these likely connections with the sun, solar hypotheses need to be taken into serious account in discussing possible methods of determining structural orientations at Giza alongside, rather than necessarily as an alternative to, sightings upon stars.

## *5.4     Equinoctial alignment strategies*

One method of identifying east is to locate the position of the rising or setting sun at the equinox. If we are going to postulate that various structures on the Giza plateau were aligned in this way then we must clarify (a) which 'equinox' we are referring to; (b) how the ancient Egyptians could have set out such alignments and what precision is likely to have been achieved; and (c) how likely is it that the equinox was significant to the ancient Egyptians?  Regarding the first question, there are at least four different 'practical' definitions of the equinox (Ruggles 1997):
1) The spatial midway point of the sun along the horizon between the winter and summer solstices, either in the spring or in the autumn.
2) The midpoint in time between two successive solstices.
3) The day on which the sun rises and sets at exactly opposite points on a level horizon.
4) The 24-hour period in which 'night' and 'day' (defined as the time between sunset and sunrise and vice versa) are equal in length.

Each of these methods will produce somewhat different results owing to varying horizon altitudes and refraction, and (in the case of method 2) the inconstant rate at which the earth travels around the sun.  For example, method 2 will typically produce declinations in the range +0.1° to +0.8° (Ruggles 1999: 54–55); which for an eastern horizon altitude of 0° corresponds to the azimuth range 88.7° to 89.5°; each increase of 1° in the altitude will increase the horizon range by some 0.7°, so that for altitude 3° the range will be 90.7° to 91.5°.

Method 1 would have involved the construction, say, of a north-south wall at a point where the horizon altitude was the same in the direction of the summer solstice and the winter solstice sunrise. The surveyors would have needed to mark the direction of both winter and summer solstice sunrise on this wall, and then divide the distance between the two into equal parts. While this is not inconceivable, there is no evidence for the existence of such a wall.

Method 3 would depend upon the use of gnomons and cords but, critically, would depend upon finding a location where the altitude of both the eastern and western horizons is close to 0°. As Table 12 shows, high horizons to the west seem to rule out the use of this method on the Giza plateau.

Method 4 presupposes the use of accurate timing devices. Star clocks date back at least to the 9th dynasty (Claggett 1995: 52) and, according to Parker (1974: 53–55), diagrams of stars inside coffin lids indicate that a system of 12 night hours was in use by at least ca. 2150 BC. Wells (1993; 1996: 36–37) argues that a previous system, recorded in the *Book of Gates* (a New Kingdom funerary text), used a 24-hour day. Be this as it may, as early as dynasty 5 there were 'watchers of the hour' priests (*Imyw-wnwt*) who kept accurate time at night by sighting and recording stellar decans, so it is not inconceivable that this was done during dynasty 4.

Method 2 might seem the most straightforward, since the 'watchers of the hour' could have determined when the equinox was near by stellar observations. Each of the 36 decans was represented by a constellation that rose for 10 days—an Egyptian decade—before a new one appeared, with 5 intercalary days completing the 365-day year. Thus the equinox would fall a quarter of a year, about 9 decades and 1–2 days, after the solstice. A major issue with this method would have been to keep track of the date of the solstice in the first place, since it would have moved forward one day in every four years.

A full analysis of the efficacy of the four methods, given the numerous cultural and technical factors that are relevant, is beyond the scope of this paper. However, we can conclude broadly, on technical grounds, that the best precision in identifying true east-west that could be achieved by any of these methods is about 0.5° (cf. Belmonte 2001: S2). As a 'rule of thumb' we would look for a range of declinations from about 0.0° to +1.0' if method 2 were used, but about –0.5° to +0.5° in other cases.

However, no records have been recovered to indicate that the ancient Egyptians were cognizant of equinoxes or that they held any importance for them.

## *5.5    Possible astronomical orientation methods used at Giza*

As we have seen, in the two cases where we have direct evidence—in the form of 'survey' postholes—of survey baselines and offsets being set up, the evidence suggests that the primary baseline was oriented north-south. The question we explore in this section is whether the orientations and relative situations of the various structures on the plateau can, in themselves, furnish any clues as to the astronomical method used to determine the orientation of the baselines. The possibilities we shall consider are:
- north-south baselines set up using observations of circumpolar stars in the north;
- north-south baselines set up using observations of shadows at local noon; and
- east-west baselines set up using observations of the rising (or setting) equinoctial sun.

We are also mindful that, in all cases, the perpendicular lines could have been set out using geometrical methods and/or set out or corrected using independent astronomical measurements.

Belmonte (2001: S1–3) has reviewed the various astronomical methods proposed in the literature to determine a cardinal orientation and concluded that only observations of a pair of circumpolar stars that simultaneously cross the meridian (as proposed by Spence 2000) could explain the extraordinary precision with which this was achieved in the case of the Khufu and Khafre pyramids. Yet Belmonte's critique of Spence's choice of stars highlights the fact that there is no clear-cut choice *for us* as to the candidate pairs of stars that best fit the orientation and dating evidence, let alone the contextual evidence regarding the possible instruments that could have been used to make the observations. This raises the concern that the mere existence of a pair of stars that fit the constraints does not necessarily provide strong proof that this was actually the method used. On the other hand, the fact that there is only a tiny (ca. 1.5′) difference in DFC between the Khufu and Khafre pyramids argues strongly that such a pair of stars *was* chosen, and that the same pair was used to orient both pyramids. Furthermore, if the builders' margins of error could have been no better than ca. 2′, as Belmonte (2001: S1) has argued, then we cannot rule out from the orientation data the possibility that the two pyramids were, in fact, aligned almost simultaneously.

Spence (2000) introduced the idea that the substantial difference in DFC between Menkaure's pyramid and the other two (some 27′ between Menkaure and Khafre, and in the opposite sense) could be due to precession. Another possibility is that a different method was used to determine Menkaure's orientation. The fact that its clockwise DFC is repeated in the burial monuments of Menkaure's immediate successors, though not with any systematic increase in the DFC (Shepseskaf +0.5°; Userkaf +0.25°; Sahura +1.5°; Neferirkara +0.25°; Belmonte 2012: 184) suggests that a distinct orientation practice was used for several pharaohs from Menkaure onwards, but that it did not involve the consistent use of the same sighting stars.

Then there are all the other structures on the plateau. How should their more disparate orientations be explained? In the light of the conclusions of the previous sections, we should explore the possibility that observations of sunrise or sunset could have been important. In Table 12 we present the principal data concerning east-west alignments of structures on the plateau, including the elevation of the structure concerned, the altitude of the horizon, and the declination. Azimuth values are quoted to the nearest 0.1°, while altitude values (measured by EN using a hand-held compass-clinometer) are quoted to the nearest 0.25°.wherever readings were checked several times, 0.5° otherwise. A precision of 0.1° is therefore justified for the declinations, which were calculated using CR's GETDEC program (Ruggles 1999: 169; still available from www.cliveruggles.net), which takes account of celestial refraction. A mean latitude value of 29.975 was assumed.

Elevations, estimated from the published elevations of control points on the GPMP grid (Goodman 2007) and Google Earth, are included to give a sense of the relative heights of the different locations on the plateau:.[15]

**Table 12: East-west alignments: summary data**

| Approx date* | Structure | Elev (m) | Alignment to east | | | | Alignment to west | | | |
|---|---|---|---|---|---|---|---|---|---|---|
| | | | Az (°) | Alt (°) | Dec (°) | Notes | Az (°) | Alt (°) | Dec (°) | Notes |
| Khufu complex | | | | | | | | | | |
| 2587 BC | Main pyramid | 59m | 90.0 | +0.25 | –0.1 | | 270.0 | +5.0 | +2.4 | |
| 2587 BC | Mortuary temple | 61m | 90.0 | +0.25 | –0.1 | a | 270.0 | +5.0 | +2.4 | a |
| 2585 BC | Queen's pyramid KQ2 | 56m | 89.7 | +0.25 | +0.1 | a | 269.7 | +5.0 | +2.2 | a |
| 2585 BC | Western cemetery | 69m | 87.7 | >+0.25 | >+1.8 | a, b, c | 267.7 | +4.5 | +0.2 | a |
| 2535 BC | Southern cemetery | 61m | 89.4 | +0.25 | +0.4 | | 269.4 | +6.0 | +2.4 | |
| 2583 BC | EC row 1 (7100) | 57m | 89.6 | +0.25 | +0.2 | | 269.6 | +6.0 | +2.6 | |
| 2583 BC | EC row 2 (7200) | 55m | 89.5 | +0.25 | +0.3 | | 269.5 | +6.0 | +2.5 | |
| 2583 BC | EC row 3 (7300) | 53m | 90.3 | +0.25 | –0.4 | | 270.3 | +6.0 | +3.2 | |
| 2583 BC | EC row 4 (7400) | 51m | 91.0 | +0.25 | –1.0 | | 271.0 | +6.0 | +3.8 | |
| Khafre complex | | | | | | | | | | |
| 2556 BC | Main pyramid | 69m | 89.9 | +0.25 | –0.1 | b | 269.9 | +15.0 | +7.3 | d |
| 2556 BC | Satellite pyramid | 65m | 90.1 | +0.25 | –0.2 | b | 270.1 | +10.0 | +5.0 | d |
| 2556 BC | Causeway (top) | 60m | 103.5 | +0.25 | –11.8 | b | — | — | — | |
| 2556 BC | Causeway (bottom) | 22m | — | — | — | | 283.3 | +8.0 | +15.4 | |
| 2556 BC | Valley temple | 20m | 90.5 | +0.25 | –0.6 | e | 270.5 | +8.0 | +4.3 | |
| 2556 BC | Sphinx temple | 24m | 90.4 | +0.25 | –0.5 | e | 270.4 | +8.0 | +4.3 | |
| Menkaure complex | | | | | | | | | | |
| 2530 BC | Main pyramid | 72m | 90.4 | +0.25 | –0.5 | | 270.4 | +6.5 | +3.5 | |
| 2530 BC | Queen's pyramid MQ2 | 67m | 90.5 | +0.25 | –0.6 | | 270.5 | +6.5 | +3.6 | |
| 2530 BC | Causeway (top) | 68m | 89.2 | +0.25 | +0.6 | b | — | — | — | |
| 2530 BC | Causeway (bottom) | 20m | — | — | — | | 269.2 | +6.0 | +2.2 | |
| 2530 BC | Valley temple | 20m | 90.2 | +0.25 | –0.3 | | 270.2 | +6.0 | +3.1 | |

---

[15] The earliest reference to the relative elevations of the three main pyramids is on a plan given by Vyse (1840: opp. 148). In all copies seen by the authors this information is illegible, but Maragioglio and Rinaldi (1967: 32–35) quote Vyse as saying that the difference in level between the bases of the pyramids of Cheops and Mycerinus is 12.68m and between Chephren and Mycerinus 2.57m. The values quoted in Table 12 are broadly consistent with this.

**Notes**
* See note 7 and Tables 2b, 4b, 6b.
a Azimuth estimated as perpendicular to the N-S direction
b The distant horizon was not visible through smog, so its altitude is calculated from topographic data.
c Depending upon the position within the cemetery, the Great Pyramid would have obscured the eastern horizon and increased the effective altitude.
d The large altitude value here is due to the presence of a manmade escarpment some 9m tall and only about 6m to the west of the pyramid, formed when the ground in the vicinity was levelled in preparation for its construction. The escarpment begins to smooth out to the south of the main pyramid (i.e. to the west of the satellite pyramid).
e The horizon is obscured by houses so its altitude is calculated from topographic data.

### 5.5.1 The pyramid complex of Khufu

It is possible that the orientation of the mastabas in the western cemetery (DFC ≈ –2.3°)—deviating from cardinality by degrees rather than minutes, in line with the earliest pyramid of Snefru, Djoser's Step Pyramid, and most of the proto-pyramids—could be explained by alignment upon a northern star or stars, possibly in the constellation *Meskhietiu*. However, an examination of Table 12 suggests other possibilities. One is that the orientations of the successive tombs built out both to the west and east of the Great Pyramid were simply determined with reference to the pyramid itself, with errors accumulating both out into the western cemetery and beyond the third row of the eastern cemetery.

Another possibility is suggested when we notice that all of the associated structures in the Khufu complex (including the southern cemetery, built after Khufu's reign), with the single exception of the mastabas in the western cemetery, are oriented in the east upon a declination between c. –1.0° and +0.5°. The western cemetery mastabas yield a declination in this range towards the west rather than the east. A factor that distinguishes the WC mastabas from all the other associated structures in the complex is that the Great Pyramid would have obscured their eastern horizon. This suggests that sunrise on the eastern horizon might have been the determining factor at all the monuments except for those within the western cemetery, where sunset on the western horizon was used instead.

The declination range does not, however, correspond to the equinoctial range: the value of –1.0° at EC row 4 falls significantly outside the –0.5° to +1.0° 'window' discussed above. Given the lack of any historical backing for an interest in the equinox, it seems clear that we must seek a different explanation. A possibility is that once the orientation of the Great Pyramid was fixed, the surveyors waited for the day to arrive when the sun was observed to rise in alignment with orientation to the east, using this same sunrise to fix the orientation of the outlying structure. In the case of the western cemetery, with the eastern horizon being obscured, they would have had to wait until sunset on that same day.

Yet another possibility is that the western cemetery orientations reflect a mixing of an older orientation tradition rather than a practical restraint. Osiris (representing the Underworld) was an important deity who had joined with the canine necropolis god, Khentimentiu, and also assumed Khentimentiu's title, "Foremost amongst the Westerners". As the sun sets in the west it begins its journey through the underworld, so it is possible that setting sun (western) tomb orientations may have been important in order to assist the deceased on their journeys into the Afterlife.

We conclude that observations of the rising or setting sun provide a viable *prima facie* explanation of the overall range and pattern of orientations encountered among Khufu's pyramids and tombs. At the same time it is a reasonable assumption that the Great Pyramid—which clearly embodied finer, more precise engineering, construction, and alignment technologies than the surrounding mastaba fields—used a distinct set of astronomical observations in order to achieve its very close alignment to the cardinal points.

### 5.5.2 The pyramid complex of Khafre

Khafre's satellite pyramid, like those of Khufu's queens, follows the orientation of the main pyramid to within about 0.3°. But unlike KQ1–3 it is situated to the south of the main pyramid. This means that whether or not Khafre's pyramid was aligned using northern stars this would certainly have been impossible in the case of the satellite, because the main pyramid would have stood in the way. It is possible that the satellite pyramid was simply aligned to the southern side of Khafre's pyramid, but with a loss of precision.

As we have noted, the north-south axes of Khafre's valley temple and the Sphinx temple are both aligned very closely to the meridian, deviating, if anything, clockwise rather than anticlockwise. As far as can be ascertained from their shorter north and south walls (excluding the passage between them, which is clearly on a different alignment), their east-west axes definitely deviate from cardinality in a clockwise sense, the deviation amounting to some 25 or 30 minutes.

The eastern declinations from these two structures do not in themselves rule out an equinoctial explanation, but the 'copying sunrise in line with the main pyramid' explanation put forward at Khufu also seems more plausible here. It fits all the associated structures in the Khafre complex (apart from the causeway), and is especially pertinent for the valley temple and Sphinx temple given that independent evidence hints at strong solar associations.[16]

On the other hand, it would appear that the north-south axis was the primary one in the case of the valley temple and the Sphinx temple. The precision of the north-south orientation of both temples—certainly within half a degree of the true meridian—implies that the principles governing their orientation were more akin to—and possibly identical to—those applying to Khafre's own pyramid, with similarly refined and precise techniques being used to determine the correct orientation.[17] This is in contrast to the more diverse practices found, for example, among the mastabas.

It is reasonable to conclude in passing that the orientation of the causeway, somewhere between the solar solstitial and equinoctial directions, was a consequence of relationships such as those just mentioned—relationships that determined the specific locations of the Khafre's mortuary and valley temples, which it connects—rather than being of significance in itself.

### 5.5.3 The pyramid complex of Menkaure

The consistency of alignment between Menkaure's pyramid, MQ2 and the valley temple is notable in itself: none of them has a DFC greater than +0.5°, and the valley temple is aligned within ~0.2° of the main pyramid despite being well separated and at a lower elevation. While the queens' pyramids could well have been oriented simply by offsetting from the southern side of Khafre's pyramid, the valley temple must have been independently aligned. It is actually closer to true cardinality than the main pyramid.

---

[16] Although no Old Kingdom texts have been discovered that refer to the Sphinx or its temple; we do know that in dynasty 18 the pharaoh Thutmose IV decided to uncover the Sphinx (most of it was buried in sand) and to reactivate the Sphinx cult (Shaw and Nicholson 1995: 277). The name given to the Sphinx at this time was *Hor-em-akhet* (Horus at the Horizon). Could there have been an association with the sun on the horizon extending back to the time of construction? The fact that the summer solstice sun sets between the pyramids of Khufu and Khafre as viewed from the paws of the Sphinx, thus forming a natural version of the *akhet* ('horizon') sign (Lehner 1997: 130), gives some support to this idea. However, Lehner's contention (ibid.: 129–130) that the equinoctial sun sets in alignment with the Sphinx temple, the south side of the Sphinx, and the south side of Khafre's pyramid is not true from the same position, since the closer horizon obscures the bottom corner of the pyramid (Belmonte, priv. comm.. 2012); in any case, as already mentioned, there is no historical evidence of any interest whatsoever in the equinox among the ancient Egyptians.

[17] Unfortunately, the available data do not permit any direct comparison between Khafre's valley temple and Khufu's, the latter being lost beneath a modern village.

As we have noted, sunrise observations are too imprecise to explain the extraordinary level of precision achieved in the case of Khufu and Khafre's pyramids, but it is conceivable that the orientations of three Menkaure structures were all fixed using the rising sun. With the easterly declinations for Menkaure's pyramids and valley temple lying between –0.6° and –0.3°, it is just possible that equinoctial sunrise could have been used consistently for all three structures, albeit with good fortune in achieving such consistent alignments. It should also be noted that Menkaure's queens' pyramids could not have been aligned using northern stars once the main core of Menkaure's pyramid had been built (and Khufu's pyramid would also have been in the way).

However, it is also possible that the main pyramid and valley temple were aligned using (the same) northern stars, while the queens' pyramids were aligned by offsetting.

But in view of our conclusions at Khufu's and Khafre's complexes, perhaps the most plausible explanation is that the orientation of Menkaure's pyramid was fixed using circumpolar stars, while the valley temple was aligned using sunrise, matching the day when the sun rose in line with the easterly orientation of the main pyramid.

It is tempting to suggest that the 0.8° anticlockwise deviation of Menkaure's causeway from true east-west might be due to its orientation having been determined by equinoctial alignment. However, a lower horizon altitude in the east would fit such an argument better: an altitude of +0.5° would yield a declination of +0.7°, altitude 0.0° would yield declination +0.4°, and altitude –0.25° would yield declination +0.2°.

### 5.5.4 Khafre's workmen's village

One other group of structures is associated with Khafre's pyramid complex: the village of the workmen who built the pyramid (Lehner 2002). This is situated to the SSE of the Sphinx, beyond the temenos wall of the ancient necropolis ('Wall of the Crow') and is currently being excavated by the Giza Plateau Mapping Project. The core of the village comprises sets of long narrow galleries running roughly north-south, not quite perpendicular to three east-west streets. Alignment data provided by the GPMP show that the galleries were aligned with azimuths between 356.6°/176.6° and 357.6°/177.6° while the streets were aligned with azimuths between 87.9°/267.9° and 88.9°/268.9°.

The elevation of the workmen's village is similar to that of the Sphinx temple and Khafre's valley temple, but it is laid out on a different orientation, skewed anticlockwise from cardinality by an average of 1.6° (streets) and 2.9° (galleries). We cannot entirely rule out the possibility that the orientation was determined by observations of the rising sun at the equinox (az = 87.9°, alt = +0.25° => dec = +1.7°; az = 88.9°, alt = +0.25° => dec = +0.8°) although the altitude of the western horizon is too great for the alignment to have been determined by observations of the setting equinoctial sun (az = 267.9°, alt = +8.0° => dec = +2.1°; az = 268.9°, alt = +8.0° => dec = +3.0°). The question of what determined the alignment of the village remains open.

### 5.5.5 The complex of Khentkawes

For completeness, we conclude this section with mention of the temple complex of Khentkawes, unique in being a female ruler during the transition between dynasties 4 and 5. Her tomb (built c. 2494 BC) is located just to the north of Menkaure's valley temple, with its causeway on the eastern side. The base of the tomb was carved out of the natural rock, and the area around it was quarried out and used for other monuments, in the same way as the area around the Sphinx had been quarried out and used to construct the Sphinx temple and Khafre's valley temple. Khentkawes' complex is unique in that it included a town (Lehner 1997: 138).

Khentkawes' tomb and entrance corridor have an orientation pattern markedly different from that of Menkaure's valley temple, causeway, and pyramid. Alignment data provided by the GPMP show that the 'east-west' alignment of the tomb, temple and town were between 82.9°/262.9° and 83.1°/263.1°. This is too far from cardinality for the orientation to have been determined by observations of the rising sun at the equinox: the altitude of the eastern horizon here, again +0.25°, yields a declination between +5.8° and +6.0°. To the west, the deviation overcompensates for the horizon altitude (+8.0°): to obtain a declination in the range –0.5° to +1.0° we would need an azimuth in the range 264.8° to 266.6°. Once again, the question of what determined the alignment remains open.

# 6 Conclusions

The pyramids of Giza have been the focus of academic, religious and spiritual attention for centuries. However, practically all of the focus has been on the pyramids of Khufu and Khafre, with scant attention paid to Menkaure's pyramid, and little or none to the orientations of the surrounding structures. It has been widely accepted that the three Giza pyramids were aligned to northern circumpolar stars, and that the slight variations between the orientations of these pyramids was due either to precession or to measurement errors.

The main aim of this paper has been to re-examine and augment the basic data relating to the orientations of the pyramids of Giza Plateau and their associated monuments. These basic data not only underlie all theories concerning how the extraordinarily precise cardinal orientation of the main pyramids was achieved, but also to enrich the contextual picture concerning the methods used when locating and orienting satellite pyramids, valley temples, other tombs, and even potentially the associated workmen's villages.

As part of our Total Station survey in 2006, we were able to collect extensive new survey data pertaining to the main pyramids, including Menkaure's, which has not been re-examined since the survey carried out by Petrie in the late 1890s. Given that the definitive baselines used by the original surveyors are not identifiable, our data emphasize the importance of considering such factors as
- variations between different levels such as casing foundations and stone courses,
- irregularities and overall concavity and convexity in individual sides and walls, and
- departures from strict parallelism in opposite sides/walls and perpendicularity in adjacent ones

before producing estimates of the actual orientations and how they were achieved at different stages, let alone a single estimate of the 'intended orientation'. These data, together with the evident variability in the results obtained by the few archaeologists from Petrie onwards who have carried out their own surveys, emphasize the false precision inherent in orientations quoted to the nearest arc second and the caution that is needed if basing detailed astronomical hypotheses on such data.

Our data suggest, for example, that there is only a very slight difference in orientation (≈0.5′) between the north-south axes of Khufu's and Khafre's pyramids, that the sides of Khafre's are more perfectly perpendicular than those of Khufu's, and that the east-west axis is closer to true cardinality in both cases.

Discussions in recent years as to how the ancient Egyptians achieved such extraordinary precision have tended to assume that the astronomical reference was the northern stars, and Spence (2000) and Belmonte (2001) have shown in particular how this could have been achieved using the simultaneous transit of two circumpolar stars. Our results do not directly challenge these ideas but do suggest some possible adjustments, e.g. in Spence's correlation between orientation and date (2000: fig. 1).

However, noting a range of evidence that suggests a strong link between the Giza monuments—and particularly the Sphinx and its temple and the Khafre Valley Temple—to solar cults, we felt it important seriously to consider the possibility that at least some of the temples ands tombs were oriented using observations of sunrise or sunset rather than circumpolar or other stars. A key question was whether observations of the equinox, despite being less precise than many postulated stellar observation strategies, could have played a major role in orienting some of the subsidiary monuments. Given the lack of any evidence that the equinox concept held any importance at all for the ancient Egyptians, we would need strong alignment evidence for this possibility to be taken seriously. Our data do not provide this.

On the other hand, our evidence has raised the possibility that while the main pyramids, and very possibly also the valley temples and the Sphinx temple, were aligned upon stellar targets to the north, many of the more minor associated structures, and particularly the many mastabas surrounding the Great Pyramid, were aligned using sunrise or (where this was not possible) sunset, using the already-established eastern axis of the main pyramid to identify the day when the sun was rising directly to the east.

In so far as our own data and tentative interpretations have lead to a consistent general picture, it is this. The observation of circumpolar stars in the north remains the most viable way of explaining the precision of cardinal orientation achieved by the pyramids of Khufu and Khafre, and the closeness of their two orientations. The stellar explanation remains the most viable at Menkaure's pyramid too, but with a significant change in the DFC due to a change in practice (now favouring a clockwise deviation from cardinality), a change in the stars used, or precession over a gap of up to 60 years—or some combination of these factors. In each complex, the satellite pyramids and valley temple (where we have the data) follow the orientation of the main pyramids to within c. 0.3°. It seems most likely that the satellite pyramids were simply aligned to the relevant side of the respective main pyramid, while the orientations of the valley temples were determined independently using similar techniques to the main pyramids.

While the orientations of the mastabas surrounding Khufu's pyramid may simply have been determined by offset, with errors accumulating at further distances from the Great Pyramid, a serious possibility is that those in the eastern cemetery were aligned primarily to the east, by reference to sunrise on the day when the sun was seen to rise in line with the northern or southern side of the Great Pyramid. In the western cemetery, where the eastern horizon was obscured by the Great Pyramid, sunset was used instead. This hypothesis is not without its difficulties: the most serious of these, perhaps, is that fact that the principal (longer) axes of the mastabas are all north-south, suggesting that this was the more important axis. That the orientations of Khafre's valley temple and the Sphinx temple may have been fixed in a similar way remains an intriguing possibility given the possible solar associations of these temples.

A complex picture involving alignment by reference to both the northern stars and sunrise to the east may seem less satisfying than one simply involving observations of northern stars, but it may fit better with the contextual evidence. For example, it is clear from the fragmentary texts and inscriptions that the ancient Egyptians had the necessary tools to align structures to the sun—gnomons, cords, merkhets, beys, plumb bobs, and stakes—and that they were also expert at land measurement, as they needed to re-stake out land annually following the inundation of the Nile. Furthermore, there is textual evidence to suggest that both stellar and sun/shadow observations could form part of the process of orienting a sacred Egyptian building. For example, one inscription reads:

> "He (the King) has built the Great Place of Re-Horakhty in conformity with the horizon bearing his disk; there the cord was stretched by His Majesty himself, having the stake in his hand with Seshat: he untied his cord with He-who-is-south-of-his-wall, in perfect work for eternity, being established on its angle by the majesty of Khnoum. He-who-makes-existence-run-its-course stood up to see its shadow, it being long in perfect fashion, wide in perfect fashion, high and low in accurate fashion, finished with work of excellent craftsmanship furnished with everything required, sprinkled with gold, decorated with colors; in appearance resembling the horizon of Re."

(Žába 1953: 60 as translated by Isler 2001: 174)

Then again, in one depiction of an alignment ceremony the king gazes at *Meskhetiu* in the presence of Thoth, the moon, the 'indicator of time' (Žába 1953: 16f). According to (Isler 1991b: 53–54), the king and Seshat could gaze at the stars and determine the correct time of year while Thoth could observe the sun's shadow using a bay with a merkhet/gnomon, with the combined observations resulting in the correct orientation.

Finally, it may well be that solar connections were also expressed in the positioning of various structures in relation to one another. It is likely that the entire Giza Plateau building project, or at least its main components, were planned before the construction of the Great Pyramid, given that
a) all four sides of each main pyramid have an unobstructed view of the cardinal points;
b) the western sides of Khufu and Khafre's pyramids align to the Mortuary Temples of Khufu and Menkaure;
c) the three pyramids form a diagonal that may have been intentionally aligned to Heliopolis; and
d) the sun at the summer solstice sets between Khufu and Khafre's pyramids, as viewed from the Sphinx; forming in the process a natural version of the *akhet* ('horizon') sign (Lehner 1997: 130).

Yet more tentative interpretations exist in the literature, an example being Wells' (1996: 30–32) conjecture that Nut (the Milky Way) swallows the sun at sunset vernal equinox, and subsequently 'gives birth' to him on winter solstice sunrise at the bifurcation of Cygnus. It is not the intention of this paper to add further speculations of this nature.

We will, however, finish with a speculation of our own. We have suggested that the orientation of some of the smaller associated structures was fixed by observing sunrise or sunset on a day when the sun was seen to rise in line with the northern or southern side of a main pyramid. Be that as it may, given the likely importance of the solar cult, it is surely probable that significance would have been attached to the dates on which the sun was seen to rise (or set) in line with the east-west axes of pyramids whose orientation had already been fixed perpendicular to a north-south axis established by stellar observations. It can hardly have escaped notice that these dates fell approximately halfway between the extreme rising and setting positions of the sun at the solstices, which came to be of great importance throughout the ancient Egyptian empire. Did this provide the motivation, and the means, by which the equinoxes were first observed, and conceived, in ancient Egypt? (cf. Shaltout *et al*. 2007: 425).

The unique concentration of monuments on the Giza Plateau contains numerous structural orientations likely to have conformed to closely related, if not completely consistent, techniques and practices. The analysis of these orientations therefore presents a unique opportunity to address key questions relating to how these orientations were determined. Subject to access restrictions, there remains plenty of scope for collecting further alignment evidence than was possible during our six days in the field. Both authors hope to have the opportunity to continue this work in the future.

## Acknowledgements

The authors are indebted to Mark Lehner for allowing us to undertake this work under the auspices of the Giza Plateau Mapping Project (GPMP) and for subsidizing CR's stay in Cairo; to their very capable field assistants Azab Hussein and Ali Fahmi; and to Supreme Council of Antiquities Inspector, Mohammed Ismail Mohammed, who facilitated their work on site. We are grateful to the School of Archaeology and Ancient History at the University of Leicester for the loan of a Leica TCR705 Total Station and CR thanks Mark Plucennik of the Distance Learning Unit for funding his travel to Egypt. Finally, we thank Juan Belmonte for reading the manuscript in advance of its submission and for his many constructive suggestions, Glen Dash for his comments, and Wilma Wetterstrom for helpful clarifications.

# Note added in proof

In a paper to be published shortly, Dash (2103) clarifies various issues to do with the casing foundation edges and corner markers at Khufu's pyramid and reanalyzes the data from Lehner and Goodman's 1984 survey, concluding that their best estimates of the orientations of the sides of the great pyramid (expressed as DFCs) are –0° 02.9′ (north), –0° 03.4′ (east), –0° 03.7′ (south) and –0° 04.6′ (west), yielding a mean DFC of –0° 03.6′. Substituting these figures in Table 8 alters the mean DFC from all five surveys as follows: north side –0° 03.0′ (unchanged); east side from –0° 04.0′ to –0° 03.9′; south side from –0° 01.7′ to –0° 02.5′; west side from –0° 04.0′ to –0° 03.5′; overall mean –0° 03.2′ (unchanged). In Table 11, the mean DFC in the north-south direction for the Khufu main pyramid is changed from –0° 04.0′ to –0° 03.7′ while that in the east-west direction is changed from –0° 02.4′ to –0° 02.7′. The conclusions that we draw on the basis of these data are unaffected.

Dash (priv. comm.) also clarifies that when Petrie found Gill's stake missing from the SW corner, he did not fill the hole with plaster (as we state in note 7) but instead drilled a small hole closer to the pyramid and used that.

We are grateful to Glen Dash for supplying a copy of his paper in advance of its publication.

*Reference*

Dash, Glen
2013        New angles on the great pyramid. *AERAgram* 13(2), 10–19.

# FIGURES

**FIG. 1.** Egyptian surveying techniques and tools known from the texts:

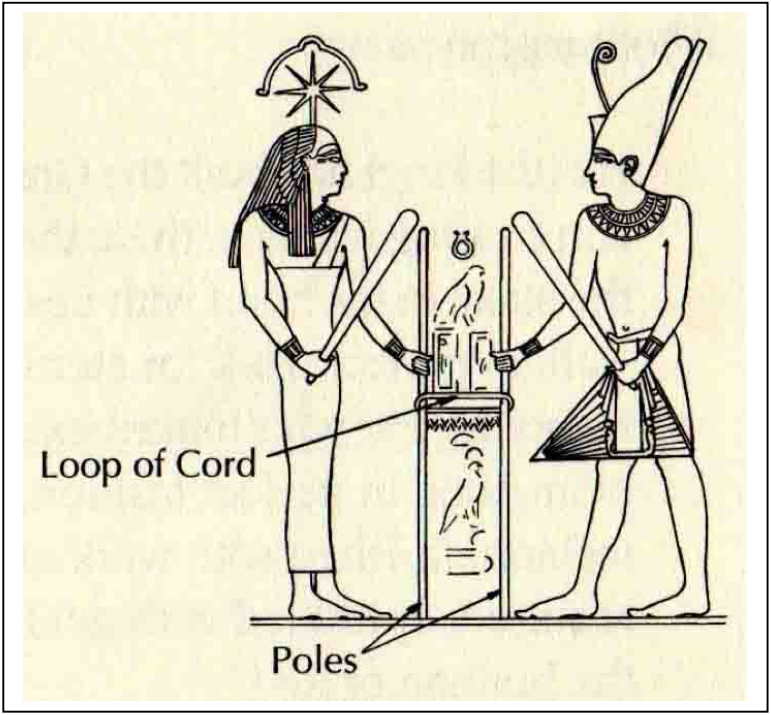

**FIG. 1(a)** 'stretching the cord' (after Isler 2001: 173)

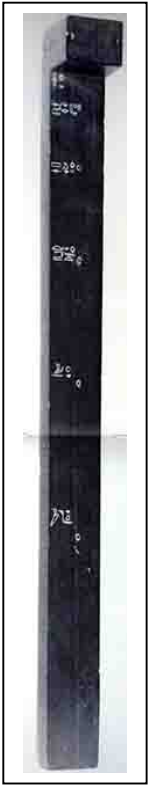

**FIG. 1(b)** the merkhet or shadow clock (after von Bomhard, 1999: 68)

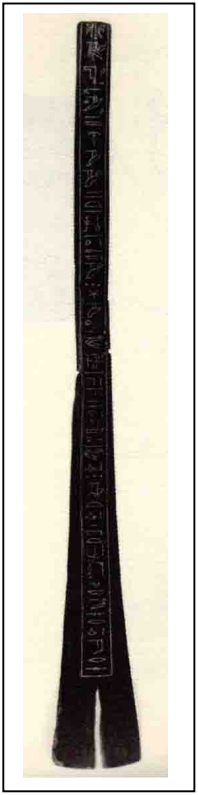

**FIG. 1(c)** the bay (after Isler 2001: 172)

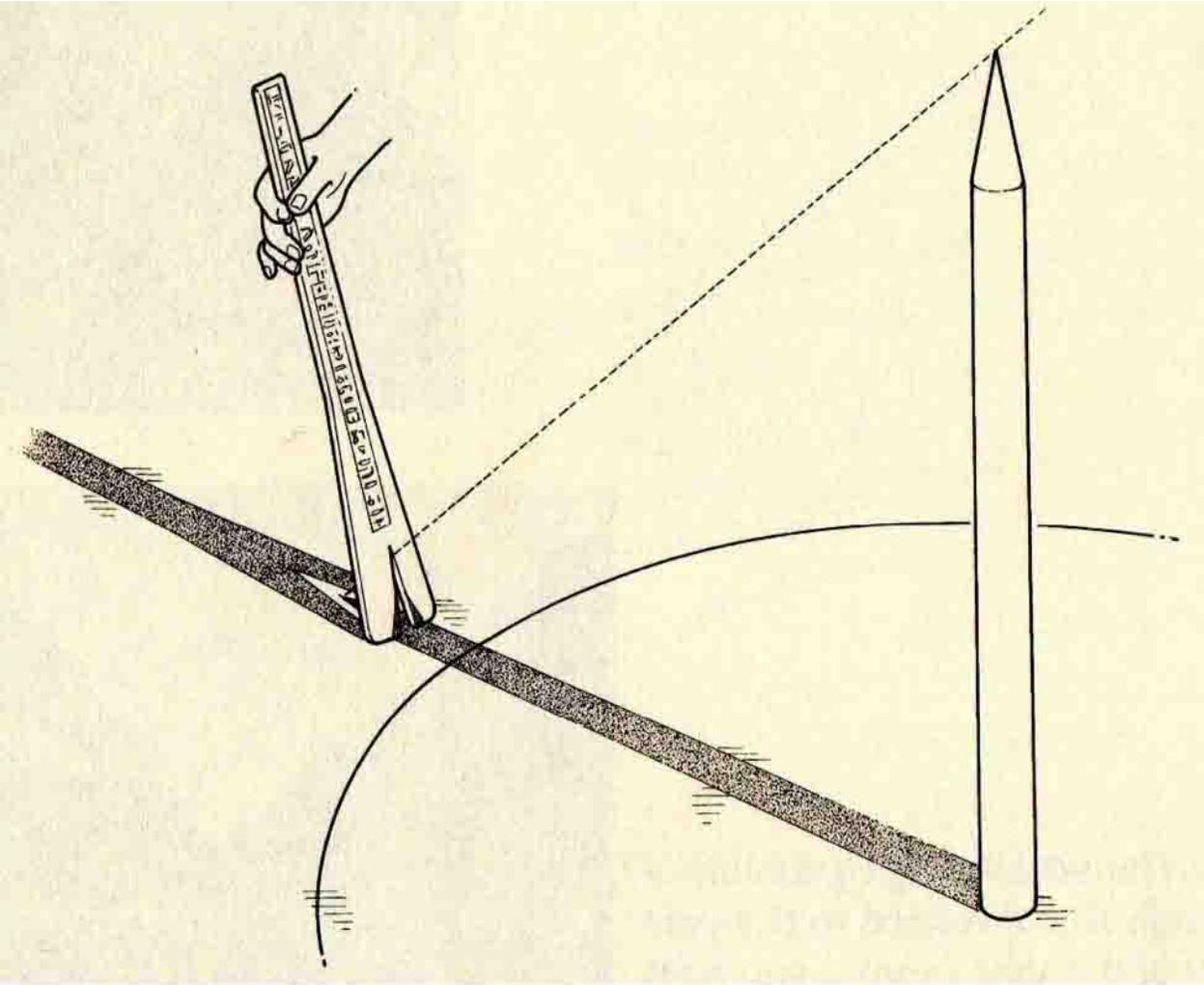

**FIG. 1(d)** use of the bay with a gnomon (after Isler 2001: 172).

**FIG. 2.** Suggested methods for determining cardinal orientations:

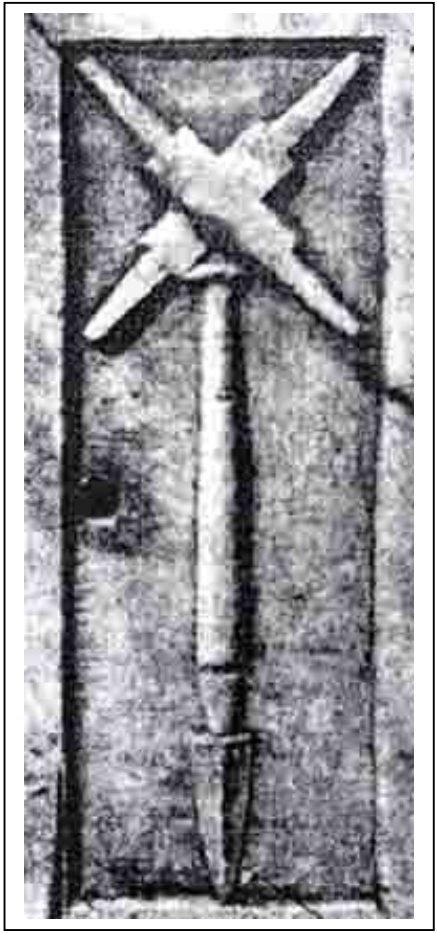

**FIG. 2(a)** A Roman stele from Pompeii depicting a Roman groma (after Miranda *et al*. 2009: fig. 3 left)

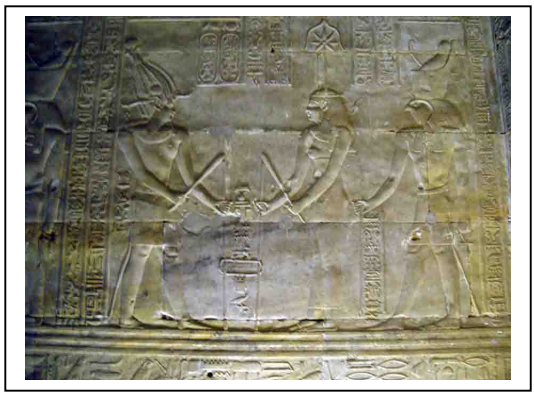

**FIG. 2(b)** Seshat's crown for comparison, as shown in the Temple of Horus at Edfu (photograph by Erin Nell)

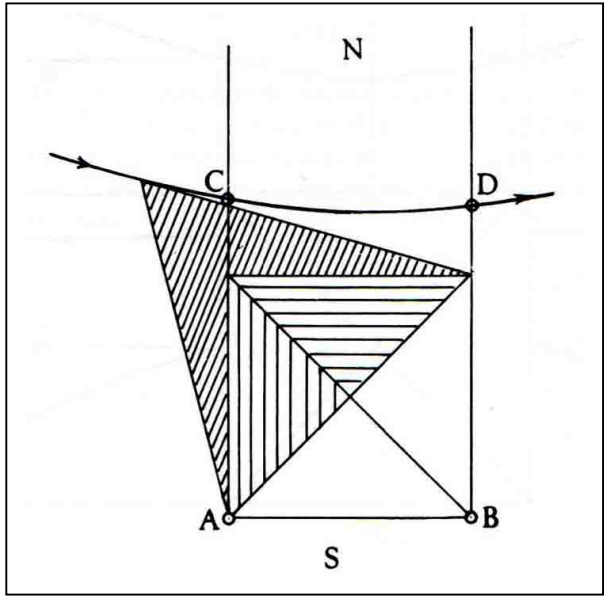

**FIG. 2(c)** Neugebauer's solar orientation suggestion (see text) (after Neugebauer 1980: 1).

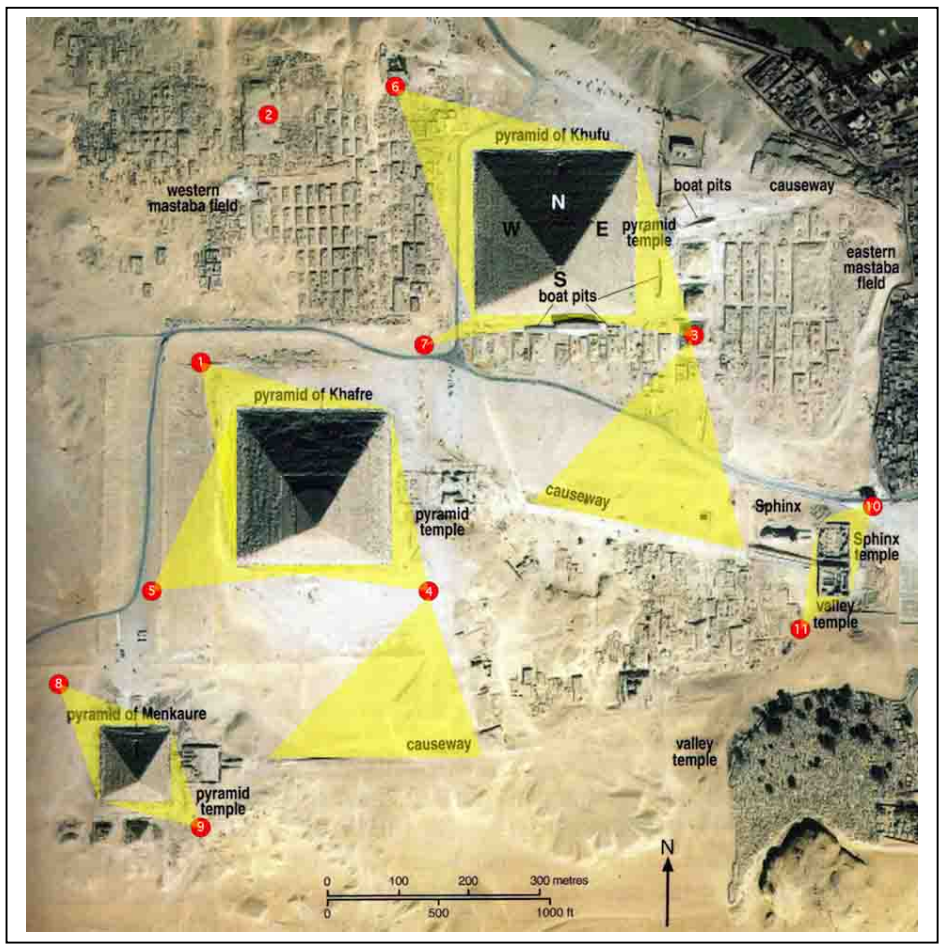

**FIG. 3.** The eleven survey stations together with an indication of the principal structures (excluding satellite pyramids and mastabas) surveyed from each of them. Underlying plan after Romer (2007: 15).

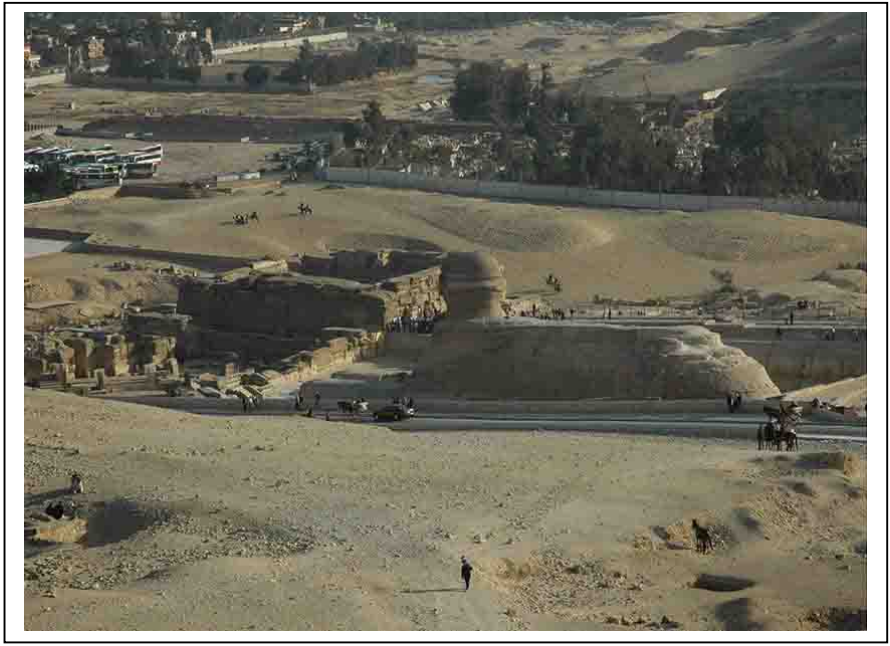

**FIG. 4.** The Sphinx and the Sphinx temple, to its left, viewed from the north. Khafre's valley temple is behind. Photograph by Clive Ruggles.

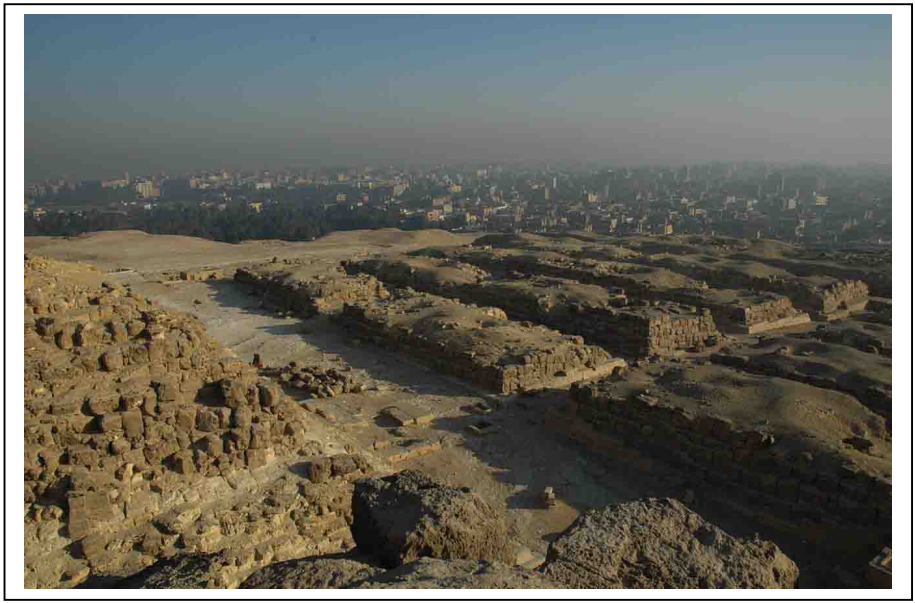

**FIG. 5(a)** Structures to the east of Khufu's pyramid, viewed northwards from the top of the southernmost Queens' pyramid (KQ3). The large structure to the far left is KQ2; the four rows of mastabas toward the right are EC–1 to 4 (G 7100 to 7400).

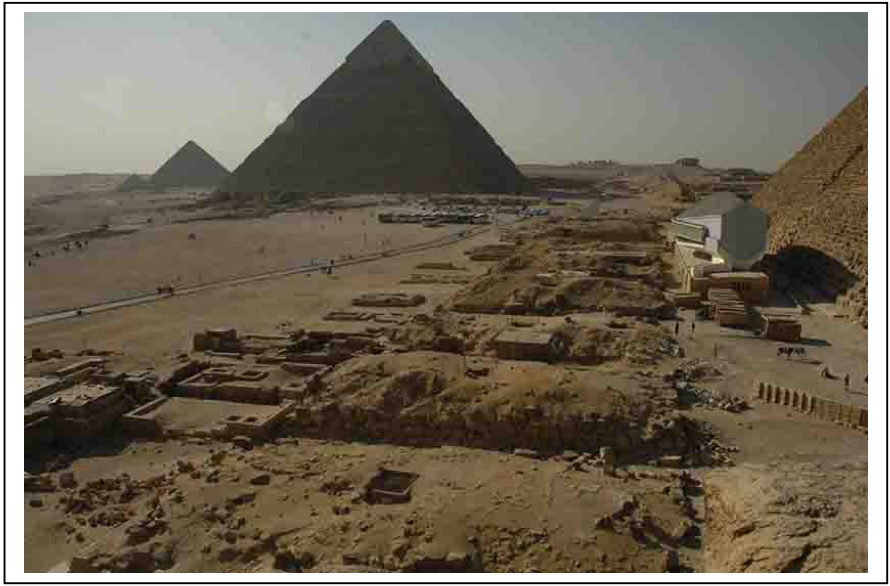

**FIG. 5(b)** Structures to the south of Khufu's pyramid, viewed westwards from the top of KQ3. In the line of mastabas, the farthest from the camera is SC–1. The base of Khufu's pyramid is seen on the right, the three pyramids in the distance being (from right to left) those of Khafre, Mekaure, and the easternmost of Menkaure's queens (MQ3). Photographs by Clive Ruggles.

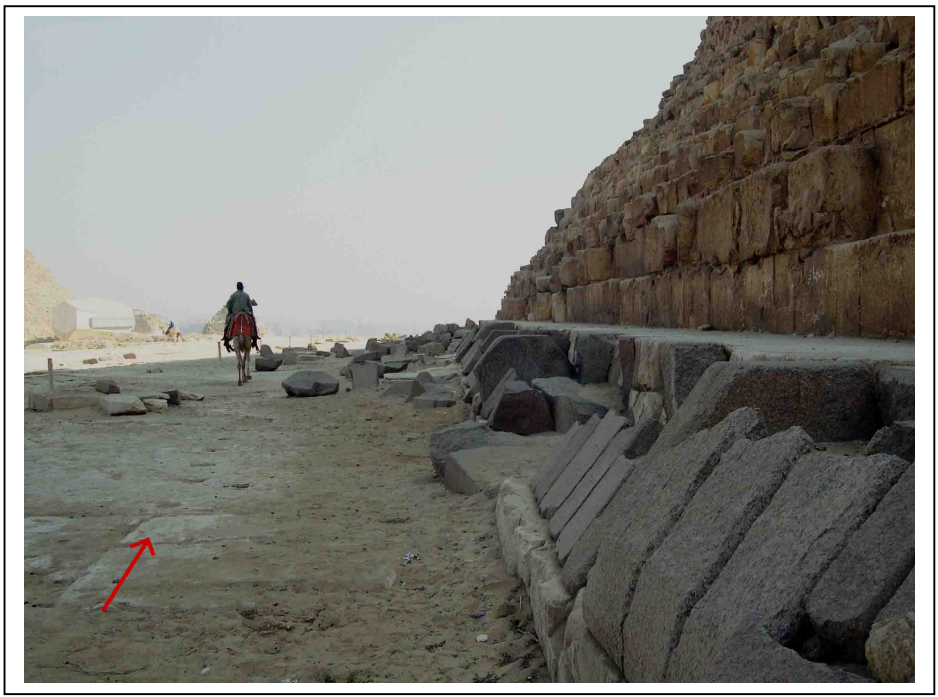

**FIG. 6.** Postholes on the east side of Khafre's pyramid. Photograph by Erin Nell.